\documentclass[twocolumn]{aastex63}

\usepackage{amsmath}
\usepackage{amssymb}
\usepackage{rotating}

\usepackage[T1]{fontenc}
\usepackage[utf8]{inputenc}

\newcommand{\ra}[4]{${#1}^{\rm h}{#2}^{\rm m}{#3}\fs{#4}$}
\newcommand{\dec}[4]{${#1}\arcdeg{#2}\arcmin{#3}\farcs{#4}$}

\newcommand\tE{t_{\rm E}}

\newcommand\piEN{\pi_{\textrm{E},N}}
\newcommand\piEE{\pi_{\textrm{E},E}}

\newcommand\piE{\pi_{\rm E}}
\newcommand\piEvec{\boldsymbol{\pi}_{\rm E}}

\received{}
\revised{}
\accepted{}

\shorttitle{Microlensing Optical Depth toward the LMC}
\shortauthors{P. Mr\'oz et al.}

\begin{document}

\title{Microlensing optical depth and event rate toward the Large Magellanic Cloud based on 20 years of OGLE observations}

\correspondingauthor{Przemek Mr\'oz}
\email{pmroz@astrouw.edu.pl}

\author[0000-0001-7016-1692]{Przemek Mr\'oz}
\affil{Astronomical Observatory, University of Warsaw, Al. Ujazdowskie 4, 00-478 Warszawa, Poland}

\author[0000-0001-5207-5619]{Andrzej Udalski}
\affil{Astronomical Observatory, University of Warsaw, Al. Ujazdowskie 4, 00-478 Warszawa, Poland}

\author[0000-0002-0548-8995]{Micha\l{} K. Szyma\'nski}
\affil{Astronomical Observatory, University of Warsaw, Al. Ujazdowskie 4, 00-478 Warszawa, Poland}

\author{Mateusz Kapusta}
\affil{Astronomical Observatory, University of Warsaw, Al. Ujazdowskie 4, 00-478 Warszawa, Poland}

\author[0000-0002-7777-0842]{Igor Soszy\'nski}
\affil{Astronomical Observatory, University of Warsaw, Al. Ujazdowskie 4, 00-478 Warszawa, Poland}

\author[0000-0002-9658-6151]{\L{}ukasz Wyrzykowski}
\affil{Astronomical Observatory, University of Warsaw, Al. Ujazdowskie 4, 00-478 Warszawa, Poland}

\author[0000-0002-2339-5899]{Pawe\l{} Pietrukowicz}
\affil{Astronomical Observatory, University of Warsaw, Al. Ujazdowskie 4, 00-478 Warszawa, Poland}

\author[0000-0003-4084-880X]{Szymon Koz\l{}owski}
\affil{Astronomical Observatory, University of Warsaw, Al. Ujazdowskie 4, 00-478 Warszawa, Poland}

\author[0000-0002-9245-6368]{Rados\l{}aw Poleski}
\affil{Astronomical Observatory, University of Warsaw, Al. Ujazdowskie 4, 00-478 Warszawa, Poland}

\author[0000-0002-2335-1730]{Jan Skowron}
\affil{Astronomical Observatory, University of Warsaw, Al. Ujazdowskie 4, 00-478 Warszawa, Poland}

\author[0000-0001-9439-604X]{Dorota Skowron}
\affil{Astronomical Observatory, University of Warsaw, Al. Ujazdowskie 4, 00-478 Warszawa, Poland}

\author[0000-0001-6364-408X]{Krzysztof Ulaczyk}
\affil{Department of Physics, University of Warwick, Coventry CV4 7 AL, UK}
\affil{Astronomical Observatory, University of Warsaw, Al. Ujazdowskie 4, 00-478 Warszawa, Poland}

\author[0000-0002-1650-1518]{Mariusz Gromadzki}
\affil{Astronomical Observatory, University of Warsaw, Al. Ujazdowskie 4, 00-478 Warszawa, Poland}

\author[0000-0002-9326-9329]{Krzysztof Rybicki}
\affil{Department of Particle Physics and Astrophysics, Weizmann Institute of Science, Rehovot 76100, Israel}
\affil{Astronomical Observatory, University of Warsaw, Al. Ujazdowskie 4, 00-478 Warszawa, Poland}

\author[0000-0002-6212-7221]{Patryk Iwanek}
\affil{Astronomical Observatory, University of Warsaw, Al. Ujazdowskie 4, 00-478 Warszawa, Poland}

\author[0000-0002-3051-274X]{Marcin Wrona}
\affil{Astronomical Observatory, University of Warsaw, Al. Ujazdowskie 4, 00-478 Warszawa, Poland}

\author[0000-0002-3218-2684]{Milena Ratajczak}
\affil{Astronomical Observatory, University of Warsaw, Al. Ujazdowskie 4, 00-478 Warszawa, Poland}

\begin{abstract}
Measurements of the microlensing optical depth and event rate toward the Large Magellanic Cloud (LMC) can be used to probe the distribution and mass function of compact objects in the direction toward that galaxy -- in the Milky Way disk, the Milky Way dark matter halo, and the LMC itself. The previous measurements, based on small statistical samples of events, found that the optical depth is an order of magnitude smaller than that expected from the entire dark matter halo in the form of compact objects. However, these previous studies were not sensitive to long-duration events with Einstein timescales longer than 2.5--3 yr, which are expected from massive ($10\mbox{--}100\,M_{\odot}$) and intermediate-mass ($10^2\mbox{--}10^5\,M_{\odot}$) black holes. Such events would have been missed by the previous studies and would not have been taken into account in calculations of the optical depth. Here, we present the analysis of nearly 20-year-long photometric monitoring of 78.7 million stars in the LMC by the Optical Gravitational Lensing Experiment (OGLE) from 2001 through 2020. We describe the observing setup, the construction of the 20-year OGLE dataset, the methods used for searching for microlensing events in the light-curve data, and the calculation of the event detection efficiency. In total, we find 16 microlensing events (thirteen using an automated pipeline and three with manual searches), all of which have timescales shorter than 1 yr. We use a sample of thirteen events to measure the microlensing optical depth toward the LMC $\tau=(0.121 \pm 0.037)\times 10^{-7}$ and the event rate $\Gamma=(0.74 \pm 0.25)\times 10^{-7}\,\mathrm{yr}^{-1}\,\mathrm{star}^{-1}$. These numbers are consistent with lensing by stars in the Milky Way disk and the LMC itself, and they demonstrate that massive and intermediate-mass black holes cannot comprise a significant fraction of the dark matter.
\end{abstract}


\section{Introduction} \label{sec:intro}

The discoveries of gravitational waves produced during mergers of massive black holes in distant galaxies by LIGO and Virgo detectors \citep{ligo_virgo_2016,ligo_virgo_2019,ligo_virgo_2021,ligo_virgo_2023} raised questions as to whether such black holes exist in the Milky Way and, if yes, whether they can be detected using other means than gravitational waves. Most black holes detected in the Milky Way using electromagnetic observations typically have masses below $15\mbox{--}20\,M_{\odot}$ \citep{corral2016}, whereas those detected by gravitational-wave detectors are on average more massive. They can reach more than $100\,M_{\odot}$ \citep{gw190521_pap1,gw190521_pap2}.

The origin of the black holes discovered by gravitational-wave detectors is a subject of vigorous debate (e.g., \citealt{costa2023} and references therein). Several authors \citep[e.g.,][]{bird2016,sasaki2016,clesse2017} proposed that some of them may be primordial black holes (PBHs), which are thought to have been produced in the early universe \citep[e.g.,][]{zeldovich1967, hawking1971, carr1974, chapline1975}. If such black holes existed in the Milky Way dark matter halo in large numbers, then they should cause long-timescale gravitational microlensing events that may be potentially detected.

Searches for massive compact objects in the Milky Way halo have been carried out since the early 1990s, following the idea put forward by \citet{paczynski1986}, who proposed monitoring the brightness of millions of stars in the Magellanic Clouds on timescales from hours to years. If the whole dark matter was composed of compact objects, for example, PBHs, then the microlensing optical depth toward the center of the Large Magellanic Cloud (LMC) should equal approximately $4.4 \times 10^{-7}$ \citep[e.g.,][]{alcock2000b,mroz2024b}. The microlensing optical depth depends on the distribution of matter along the line of sight, and it can be interpreted as the probability that a given source star will lie within the Einstein radius of some microlens along the line of sight and thus be magnified by at least $A>1.34$.

The subsequent studies of microlensing events toward the LMC found that the optical depth is substantially smaller than the value expected from the dark matter halo. \citet{alcock2000b} estimated $\tau = 1.2^{+0.4}_{-0.3} \times 10^{-7}$ after the analysis of 5.7~yr of time-series photometric data for 11.9 million stars observed by the MACHO project. A reanalysis of the MACHO data by \citet{bennett2005} found a slightly lower value of $\tau = (1.0 \pm 0.3) \times 10^{-7}$. However, independent studies by EROS \citep{tisserand2007} and the Optical Gravitational Lensing Experiment \citep[OGLE;][]{wyrzykowski2009,wyrzykowski2011} did not confirm the MACHO findings and found that the optical depth toward the LMC is an order of magnitude lower. Moreover, it was realized that the measurements by EROS and OGLE are consistent with the self-lensing phenomenon (background stars are lensed by foreground objects in the LMC; \citealt{sahu1994}), leaving little room for compact objects in the Milky Way halo \citep[e.g.,][]{calchi_novati_2009,calchi_novati_2011}.

The EROS-2 survey used the MarLy 1-meter telescope at La Silla Observatory, Chile, and monitored the LMC between 1996 and 2003. \citet{tisserand2007} detected only one microlensing event in a sample of 7 million stars observed by EROS for 6.7~yr and inferred an upper limit on $\tau < 0.36 \times 10^{-7}$.

The OGLE project uses the 1.3-m Warsaw Telescope located at Las Campanas Observatory, Chile. The observations of the LMC by OGLE started during the second phase of the survey (OGLE-II; \citealt{udalski1997}) in the years 1996--2000. During that time, 21 fields were observed that covered a central part of the LMC (with a total area of 4.72\,deg$^2$). A $2048 \times 2048$ camera with a pixel size of 0.417 arcsec\,pixel$^{-1}$ was operated in a drift-scan mode, allowing for observations of a field $2048 \times 8192$ pixels wide in one exposure. The OGLE-II data were analyzed by \citet{wyrzykowski2009}, who reported the detection of only two candidates for microlensing events and measured the microlensing optical depth of $(0.43 \pm 0.33) \times 10^{-7}$.

A larger area of 40\,deg$^2$ toward the LMC was observed during the third phase of the survey (OGLE-III; \citealt{udalski2003}) during the years 2001--2009. The project operated a mosaic camera comprising eight $2048 \times 4096$ CCD detectors, with a pixel scale of 0.26 arcsec\,pixel$^{-1}$. \citet{wyrzykowski2011} found four candidate microlensing events in the OGLE-III data. They estimated the microlensing optical depth of $(0.16 \pm 0.12) \times 10^{-7}$ toward the LMC.

From the observational point of view, the microlensing optical depth can be estimated from
\begin{equation}
\tau = \frac{\pi}{2N_{\rm s}\Delta T}\sum_{i}\frac{t_{\mathrm{E},i}}{\varepsilon(t_{\mathrm{E},i})},
\label{eq:tau}
\end{equation}
where $t_{\mathrm{E},i}$ is the Einstein timescale of the $i$th event, $\varepsilon(t_{\mathrm{E},i})$ is the detection efficiency of the event, $\Delta T$ is the duration of observations, and $N_{\rm s}$ is the number of source stars observed in the experiment. The summation is performed over all events detected.

Because of the limited duration of the MACHO, EROS, OGLE-II, and OGLE-III surveys (from 5 to 9~yr), these experiments were not sensitive to very long-duration microlensing events that are expected from massive ($10\mbox{--}100\,M_{\odot}$) and intermediate-mass ($10^2\mbox{--}10^5\,M_{\odot}$) black holes ($t_{\rm E} \approx 1.7\,\mathrm{yr}\sqrt{M/100\,M_{\odot}}$, where $M$ is the mass of the lens). An inspection of detection efficiency curves published by \citet{alcock2000b}, \citet{tisserand2007}, and \citet{wyrzykowski2009,wyrzykowski2011} reveals zero efficiency at $t_{\rm E} \gtrsim 2.5\mbox{--}3\,\mathrm{yr}$. If black holes with masses in the range $10\mbox{--}1000\,M_{\odot}$ existed in a large number in the Milky Way halo, they would have been missed by these experiments and would not have been taken into account in calculations of the optical depth.

In this paper, we analyzed the data collected during the fourth phase of the OGLE project (OGLE-IV; \citealt{udalski2015}) during the years 2010--2020. Because the OGLE-III and OGLE-IV surveys had a similar setup, we combined the two data sets to create homogeneous, nearly 20 yr long light curves for about 33 million objects. An additional 29 million objects were observed only during the OGLE-IV phase. We used the combined OGLE-III/OGLE-IV data to search for microlensing events with timescales as long as $\sim 10$ yr, filling the gap in the sensitivity of the previous surveys. Our main goal was to determine whether long-timescale events can significantly contribute to the optical depth toward the LMC.

In Section~\ref{sec:data}, we describe the new combined OGLE-III/OGLE-IV data set used in the analysis. The selection of microlensing events and their properties are discussed in Sections~\ref{sec:events} and \ref{sec:detected}. In Section~\ref{sec:stars}, we estimate the number of source stars observed in our experiment. Section~\ref{sec:eff} describes the simulations used to measure the event detection efficiency. The measurements of the optical depth and event rate toward the LMC are presented and discussed in Sections~\ref{sec:tau} and \ref{sec:discussion}. The calculation of the constraints on the frequency of PBHs in dark matter is presented in detail in a companion paper \citep{mroz2024b}.

\section{Data}
\label{sec:data}

\subsection{OGLE-IV Data}

The fourth phase of the OGLE project (OGLE-IV; \citealt{udalski2015}) was commenced in 2010 March. However, the regular observations started a few months later, in 2010 June. We analyzed the data collected from 2010 June 29 ($\mathrm{JD} = 2455378$) to 2020 March 15 ($\mathrm{JD} = 2458924$)\footnote{The OGLE-IV operations were forced to stop in 2020 March due to the COVID-19 pandemic. The observations were resumed in 2022 August.}, in total nearly 10~yr of the time-series data. 

The observations were taken with a mosaic CCD camera containing thirty-two $2048 \times 4102$ detectors with a pixel scale of 0.26 arcsec\,pixel$^{-1}$, resulting in a total field of view of about $1.4\,\mathrm{deg}^2$. A detailed description of this instrument was presented by \citet{udalski2015}.

\begin{figure*}
\centering
\includegraphics[width=\textwidth]{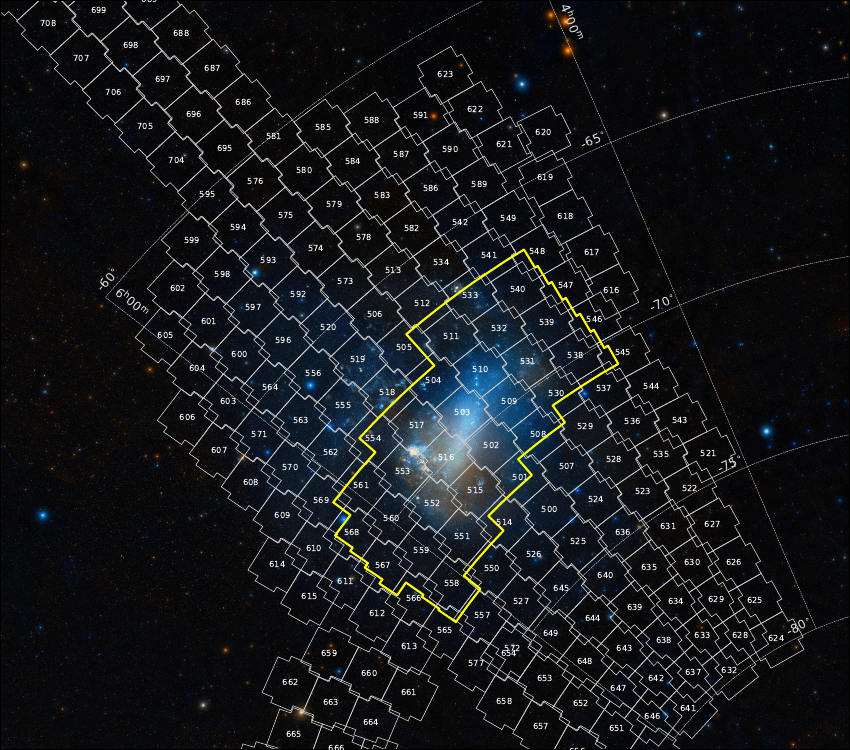}
\caption{OGLE-IV fields toward the Large Magellanic Cloud (white polygons); numbers in the polygons indicate the field number. The yellow line marks the region observed during the OGLE-III phase. The background image is reproduced with permission from WikiSky.org.}
\label{fig:fields}
\end{figure*}

We analyzed observations of 213 fields covering nearly 300\,deg$^2$ around the LMC. The location of the fields in the sky is presented in Figure~\ref{fig:fields}. Table~\ref{tab:fields} includes the basic information about the analyzed fields, including their equatorial coordinates.

The vast majority of observations were taken through the $I$-band filter, closely resembling that of the Cousins photometric system. The typical exposure time was 150\,s, providing photometry in the range of $13 \lesssim I \lesssim 21.7$\,mag. Each OGLE-IV field was observed in the $I$ band from 121 to 911 times (with a median number of 368 epochs), with a typical cadence of 3--10 days. Some additional observations (from 5 to 313 with a median of 16, depending on the field) were collected in the $V$-band filter (similar to that of the standard Johnson system) to characterize the objects.

The photometry was performed using the difference image analysis (DIA) technique \citep{tomaney1996, alard1998, wozniak2000}. For each field observed by OGLE-IV, a reference image was created by stacking several (typically 3--8) high-quality, low-seeing frames. The reference image was then subtracted from the incoming frames, and the photometry was performed on the subtracted images. Variable sources that were detected on subtracted images were flagged. See \citet{udalski2015} for a detailed description of the image reduction pipeline, photometric calibration, and astrometric transformations. 

Reference images for the analyzed fields were constructed from the images taken between 2010 and 2012. Because the dispersion of proper motions of stars in the LMC is typically smaller than 0.05\,mas\,yr$^{-1}$ in one direction \citep{mroz2024b}, the positions of stars changed by less than 0.002 OGLE pixel, even after 10~yr, enabling us to achieve high-quality photometric observations throughout the entire analyzed period. This may not be the case for foreground, high proper-motion stars, whose light curves may exhibit systematic trends. 

\subsection{Combined OGLE-III/OGLE-IV Data}
\label{sec:newdata}

The observing setup of the OGLE-IV survey was similar to that of OGLE-III. Therefore, we explored the possibility of combining the two data sets to create homogeneous 20 yr long light curves. A naive approach would involve cross-matching stars detected on OGLE-III and OGLE-IV reference images and then combining their light curves. However, because the reference images differ, the OGLE-III and OGLE-IV light curves may be systematically shifted, especially in the most crowded fields. This situation can occur when two stars are resolved in OGLE-IV but may not be resolved in OGLE-III (or vice versa).

Instead, we reduced the OGLE-III images using the OGLE DIA pipeline \citep{wozniak2000} using the OGLE-IV reference images. As discussed above, the dispersion of proper motions of stars in the LMC is small enough that decent subtractions were possible even for the earliest OGLE-III frames. A significant difficulty in performing the DIA photometry was connected to the fact that OGLE-III and OGLE-IV fields were different, so the images had to be split into smaller chunks, rotated, shifted, and resampled before performing the DIA photometry. In rare cases, when the overlap between OGLE-III and OGLE-IV images was small, the OGLE-III data may be missing for individual objects.

Table~\ref{tab:fields} presents a list of OGLE-IV fields together with the number of stars observed during both OGLE-III and OGLE-IV (that is, from 2001 to 2020) and during only OGLE-IV (from 2010 to 2020). The OGLE-III data were available for 33 million stars, and an additional 29 million objects were observed only during the OGLE-IV phase. Note that this is not equivalent to the number of microlensing source stars that enter the optical depth calculations (Equation~(\ref{eq:tau})). OGLE-III fields were observed from 385 to 637 times in the $I$ band. However, some stars may be located in overlapping fields, and the available number of epochs may be larger. The number of epochs, separately for OGLE-III and OGLE-IV phases, is shown in the fourth and fifth columns of Table~\ref{tab:fields}.

Although the observing setups of OGLE-III and OGLE-IV were very similar, different filters were used \citep{udalski2003,udalski1997,udalski2015}. The transmission curves of $I$-band filters (both of which were very similar to that of the standard Cousins filter) were slightly different. We thus found a color-dependent shift between mean magnitudes in the new OGLE-III data set and the OGLE-IV light curves. This shift was independent of the field. The median shift for stars in the color range $-0.171 \leq V-I \leq 2.602$ is presented in Table~\ref{tab:color_shift}. 

\begin{figure*}
\centering
\includegraphics[width=0.6\textwidth]{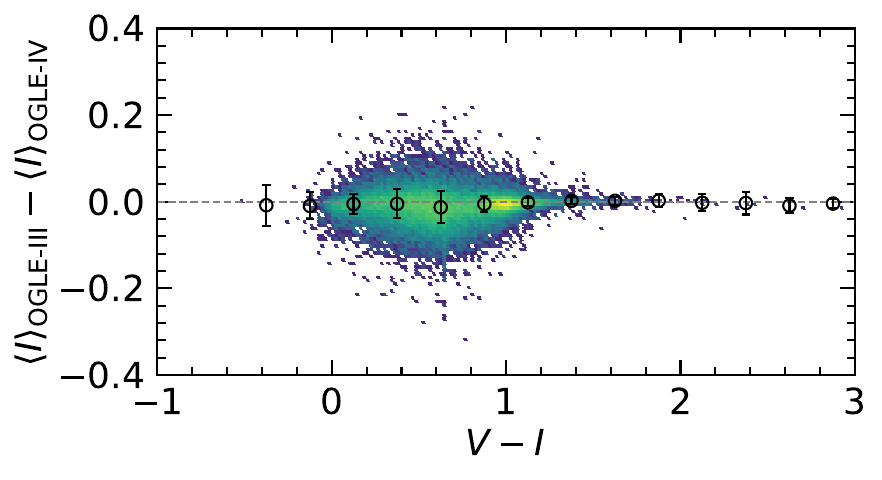}\\
\includegraphics[width=0.6\textwidth]{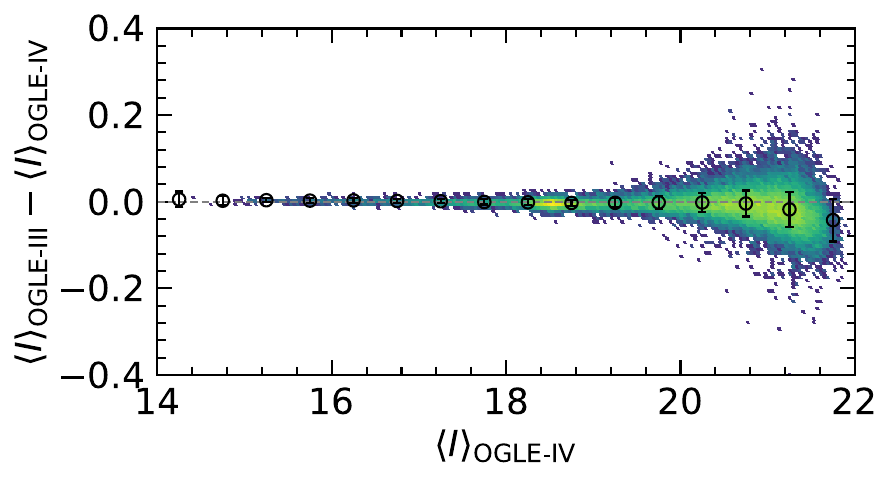}
\caption{Comparison between the new OGLE-III and OGLE-IV mean magnitudes for one of the analyzed fields (LMC502.11), as a function of color (upper diagram) and mean magnitude (lower diagram). The black data points represent the median difference calculated in several color (magnitude) bins. The error bars are calculated as the median absolute deviation divided by $\sqrt{2}\mathrm{erf}^{-1}(\frac{1}{2})$. The color scale is logarithmic.}
\label{fig:comparison}
\end{figure*}

We corrected all new OGLE-III photometric measurements for all analyzed stars by adding a constant magnitude shift that depends on the mean color of the star. The correction was found from the cubic spline interpolation of data presented in Table~\ref{tab:color_shift}. For stars with colors lying outside the interpolation range, we estimated the correction using the linear extrapolation of the trend. Figure~\ref{fig:comparison} presents the comparison between the final new OGLE-III and OGLE-IV mean magnitudes for one of the analyzed fields (LMC502.11) as a function of color (upper diagram) and mean magnitude (lower diagram). The mean difference between OGLE-III and OGLE-IV mean magnitudes is consistent with zero in the entire analyzed range.

\begin{figure*}
\centering
\includegraphics[width=0.6\textwidth]{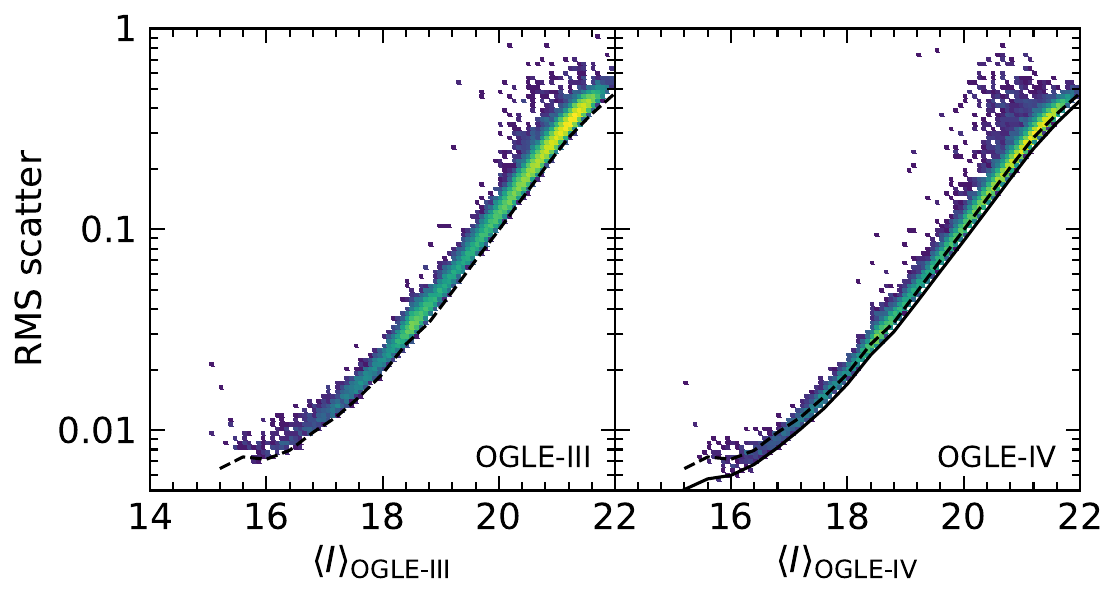}
\caption{Rms scatter in the light curves as a function of mean magnitude (for stars in the field LMC502.11). Left: OGLE-III data (2001--2009). Right: OGLE-IV data (2010--2020). The black dashed (solid) line marks the lower envelope of the distribution for OGLE-III (OGLE-IV) data. The color scale is logarithmic.}
\label{fig:noise}
\end{figure*}

Figure~\ref{fig:noise} shows the comparison between the rms scatter in the new OGLE-III and OGLE-IV light curves for one of the analyzed fields (LMC502.11). The black dashed and solid lines mark the lower envelope of the distributions for the new OGLE-III and OGLE-IV data sets, respectively. The new OGLE-III light curves are slightly more noisy than the OGLE-IV light curves; their mean scatter is larger by $\approx 12\%$.

About 5.5 million objects in the central LMC fields were observed in the second phase of the OGLE survey (OGLE-II) in 1996--2000. Although including the OGLE-II data would extend their light curves to 24~yr, boosting the sensitivity to long-timescale microlensing events, we decided not to do so. The OGLE-II camera had a different pixel size than the OGLE-III and OGLE-IV cameras, making it nearly impossible to obtain homogeneous DIA photometry across all three survey phases. However, whenever the OGLE-II data were available, we used them to check the archival photometric behavior of the analyzed events.

\subsection{Correction of Error Bars}
\label{sec:errors}

The error bars that are returned by the DIA pipeline are known to be underestimated \citep[e.g.,][]{skowron2016}. We therefore rescaled the error bars using the formula $\delta m_{\mathrm{new}} = \sqrt{(\gamma\delta m_{\mathrm{old}})^2 + \varepsilon^2}$, where $\gamma$ and $\varepsilon$ are coefficients depending on the field and filter. We closely followed the procedures outlined by \citet{skowron2016} to measure the values of $\gamma$ and $\varepsilon$. In short, we calculated the rms scatter of constant stars in a given field as a function of magnitude. We then found $\gamma$ and $\varepsilon$ for which the rescaled uncertainties follow the statistical behavior of nearby constant stars. Because the noise properties of the new OGLE-III and OGLE-IV light curves are slightly different (Figure~\ref{fig:noise}), we calculated and applied the coefficients separately for OGLE-III and OGLE-IV light curves. The best-fit coefficients are reported in Table~\ref{tab:errors}.

\section{Search for Events}
\label{sec:events}

We searched for microlensing events using the database of 62 million light curves of objects detected on OGLE-IV reference images. The OGLE-III data were available for 33 million stars and were extracted using the procedure described in Section~\ref{sec:data}. The uncertainties reported by the photometric pipeline were corrected as described in Section~\ref{sec:errors}, and then, magnitudes were transformed into flux units.

Our search procedure was similar to that developed by \citet{mroz2017}, who analyzed OGLE-IV light curves of stars in high-cadence Galactic bulge fields. It consisted of three steps, which are summarized in Table~\ref{tab:cuts}. First, objects showing any brightening with respect to the flat part of the light curve were selected. We placed a window of length $W$ on a light curve and calculated the mean flux ($F_{\rm base}$) and standard deviation ($\sigma_{\rm base}$) using data points outside the window (after removing $4\sigma$ outliers). We also calculated $\chi^2_{\rm out} = \sum_i(F_i - F_{\rm base})^2 / \sigma_i^2$, where $(F_i, \sigma_i)$ is the flux and its uncertainty of the $i$th data point outside the window.
We then searched for at least $n_{\rm bump}$ consecutive data points brighter than $F_{\rm base} + 3\sigma_{\rm base}$ within the window (which we called a `bump'). If the bump was identified, we calculated several quantities, including its amplitude $\Delta m$, duration $\Delta t$, and significance $\chi_{3+}=\sum_i(F_i-F_{\rm base})/\sigma_i$, where the summation was performed over all data points within the bump. We also counted the number of epochs $n_{\rm DIA}$ for which the DIA pipeline detected the object on subtracted images. Subsequently, the window was shifted by $S$ days, and the entire procedure was repeated until the end of the light curve was reached. A bump with the largest value of $\chi_{3+}$ was finally selected. 

For objects with both OGLE-III and OGLE-IV data available, we ran the algorithm twice using $(W,S) = (6000, 500)$ and $(3000,500)$ days. We also ran the algorithm separately on OGLE-III and OGLE-IV light curves with $(W,S) = (1500, 300)$ days. This procedure enabled us to select events lasting several years and those with timescales of several days. For the next step, we selected events fulfilling the following criteria: (a) $n_{\rm bump} \geq 5$, (b) $n_{\rm DIA} \geq 3$, (c) $\chi_{3+} \geq 32$, (d) $\Delta m \geq 0.1$ mag. If the duration of the bump was shorter than 1000 days, we also required $\chi^2_{\rm out}/\mathrm{dof} \leq 2$ (that is, no significant variability outside of the window). For bumps longer than 100\,days, we required their amplitude to be at least 0.4\,mag to minimize the contamination from stars varying on long timescales.  This simple procedure allowed us to reduce the number of analyzed light curves from over 62 million to 56,000.

In the next step, we further reduced the number of candidate events by eliminating obvious nonmicrolensing light curves. The vast majority of removed objects were ``blue bumper'' stars, which are bright, and blue main-sequence stars located in the optical color--magnitude diagrams in the region defined by $(V-I)_0 \leq 0.5$ and $I_0 \leq 19.5$. Outbursts of ``blue bumper'' stars are known to mimic light curves of gravitational microlensing events \citep[e.g.,][]{alcock1997c}. Because these outbursts may be repeating, many objects can be removed by requiring that there was only one brightening in the analyzed light curve. In addition to removing objects with multiple peaks, we removed all stars brighter than $I_0 \leq 19.5$ and bluer than $(V-I)_0 \leq 0.5$ from our sample. We dereddened the observed colors and magnitudes using the reddening maps of \citet{skowron2021}. Finally, we removed variable objects in the vicinity ($30'$) of the supernova SN~1987A remnant. These are known to be artifacts caused by the moving light echo from the supernova \citep[e.g.,][]{tisserand2007,wyrzykowski2011}. Over 7000 objects were removed in this step.

Finally, a microlensing point-source point-lens (PSPL) model was fitted to the data, and several goodness-of-the-fit statistics were evaluated. The model is described by a formula
\begin{equation}
F_i = F_0 \left[ 1 + f_{\rm s} \left(A(t_i; t_0, \tE, u_0) -1\right)\right],
\end{equation}
where $f_{\rm s} = F_{\rm s} / (F_{\rm s} + F_{\rm b})$ is the dimensionless blending parameter, and $F_0$ is the baseline flux. Here, $F_{\rm s}$ and $F_{\rm b}$ are the source star flux (which is magnified during the event), and the additional blend flux (which remains constant throughout the event, and may come from unrelated stars blended with the source and/or the lens itself and/or companions to the source and lens), respectively. The magnification was evaluated using the \citet{einstein1936} formula
\begin{equation}
A(t_i; t_0, \tE, u_0) = \frac{u^2+2}{u\sqrt{u^2+4}},
\label{eq:magnification}
\end{equation}
where $u = \sqrt{\left((t_i-t_0)/\tE\right)^2+u_0^2}$ \citep{paczynski1986}, $t_0$ is the time of the smallest lens--source separation, $u_0$ is the minimal lens--source separation (in Einstein radius units), and $\tE$ is the Einstein radius crossing timescale. 

The best-fit parameters were found by minimizing the function
\begin{equation}
\chi^2 = \sum_{i=1}^N \left(\frac{F_i - F_0 \left[ 1 + f_{\rm s} \left(A(t_i; t_0, \tE, u_0) -1\right)\right]}{\sigma_i}\right)^2
\label{eq:chi2}
\end{equation}
using the Levenberg--Marquardt algorithm as implemented in the GNU Scientific Library,\footnote{https://www.gnu.org/software/gsl/} where $N$ is the number of data points. The starting values were calculated based on the properties (time of the peak, duration, amplitude) of the bump. The algorithm yielded the best-fit parameters and their covariance matrix. We fitted two types of models: one with the blend flux set to zero ($F_{\rm b}=0$) (a ``four-parameter'' model), and one with the blend flux allowed to vary (a ``five-parameter'' model). We used the five-parameter model as a default model. However, if the five-parameter fit did not converge or it yielded an unphysical value of the blending parameter ($f_{\rm s} - \sigma(f_{\rm s}) \geq 1$), we used the four-parameter fit instead.

We calculated $\chi^2$ of the best-fit model for the entire data set, as well as $\chi^2_{\tE}$ and $\chi^2_{\rm bump}$ for the data points within $\tE$ of the peak and within the bump, respectively. We required that the model described the data well, i.e., all three statistics $\chi^2/\mathrm{d.o.f.} \leq 2$. Additionally, we required the best-fit value $u_0 \leq 1$, the best-fit $t_0$ should be within the time range covered by the data, the source magnitude should be brighter than $I_{\rm s}=22$ mag, and the Einstein timescale to be reasonably well measured, that is, $\sigma(\tE) / \tE \leq 1$. Finally, we fitted a straight line to the data and calculated $\chi^2_{\rm line}$ of the best-fit model. We required that the microlensing model should describe the data better than the straight line, that is, $\chi^2_{\rm line} - \chi^2 \geq 250 \chi^2 / \mathrm{d.o.f.}$ 

We did not take into account the microlensing parallax effect \citep{gould1992} when performing the search for the events. The microlensing parallax induces a subtle deviation from the standard PSPL model in the light curves of long-timescale events ($\tE \gtrsim 60\mbox{--}100$\,days) due to the orbital motion of the Earth. \citet{blaineau2020} demonstrated that neglecting the parallax in the fits affects marginally the selection of long-timescale microlensing events. Because the microlensing parallax $\piE$ scales inversely proportional with the square root of the lens mass $M$ (that is, $\piE \propto M^{-1/2}$), we expect that the parallaxes of events due to high-mass lenses would be small. For the Milky Way and LMC dark matter halo models presented by \citet{mroz2024b}, the mean microlensing parallaxes scale as $\piE \approx 0.1 / \sqrt{M/M_{\odot}}$ and $\piE \approx 0.02 / \sqrt{M/M_{\odot}}$, respectively. On the other hand, the microlensing parallax can be neglected for short-timescale events because the relative lens--source trajectory can be accurately approximated as a straight line.

Only 13 objects passed all selection cuts, forming the final statistical sample of events analyzed in this and companion papers. All events are listed in Table~\ref{tab:events}, together with their equatorial coordinates, mean $I$-band magnitude, and $V-I$ color in the baseline. Events OGLE-LMC-04 and OGLE-LMC-05 were previously identified by \citet{wyrzykowski2011} in the OGLE-III data. The remaining 11 events were new discoveries; we named them OGLE-LMC-NN, where NN runs from 07 through 17. The event OGLE-LMC-15 was announced in 2018 September by the Microlensing Observations in Astrophysics \citep{bond2001} survey as MOA-2018-LMC-003.

In addition to automated searches, we manually inspected the light curves of all 56,358 objects that passed the ``cut 1'' criteria. We identified three additional events. OGLE-LMC-06 was previously found by \citet{wyrzykowski2011} as a binary-lens event. Because the standard PSPL model poorly fits the data, this event was rejected by the ``cut 3'' criteria. The best-fit source magnitude of OGLE-LMC-18 placed it below our limit ($I_{\rm s} = 22$\,mag), while OGLE-LMC-19 was located in the region occupied by ``blue bumper'' stars in the color--magnitude diagram. The light curve of the latter object is symmetric, has high amplitude, and can be well fitted by a standard PSPL microlensing model, so there is no reason to suspect this is a false positive. The three events identified in by-eye searches were not included in the following statistical analysis.

\section{Detected Events}
\label{sec:detected}

Once we identified all microlensing events in our sample, we used the Markov Chain Monte Carlo (MCMC) methods to refine the best-fit parameters and their uncertainties. We assumed flat (uniform) priors on all parameters except the prior on the blend flux:
\begin{equation}
\mathcal{L}_{\rm prior} =
\begin{cases}
    1 & \text{if } F_{\rm b} \geq 0,\\
    \exp\left(-\frac{F_{\rm b}^2}{2\sigma_{\rm b}^2}\right) &  \text{if } F_{\rm b} < 0,
\end{cases}
\label{eq:prior}
\end{equation}
where $\sigma_{\rm b}$ is the flux corresponding to a $I=20.5$ star. This prior enabled us to avoid solutions with unphysically large values of negative blend flux. We ran the MCMC using the code \textsc{Emcee} by \citet{foreman2013}, providing us with posterior distributions for model parameters for all events. The modeling results, including the median and the 68\% confidence range of the marginalized posterior distributions, are summarized in Table~\ref{tab:params}. (Table~\ref{tab:params} does not contain OGLE-LMC-06, because its light curve cannot be adequately described by a PSPL model.) The best-fit models are plotted in Figure~\ref{fig:light_curves}.

In addition to standard PSPL models, we also fitted models incorporating the annual microlensing parallax effect. The modeling was performed in the geocentric frame \citep{gould2004}, with the two additional parameters $(\piEN, \piEE)$ (north and east components of the microlensing parallax vector $\piEvec$) describing the light curve. The formula for magnification is identical to that in Equation~(\ref{eq:magnification}). However, the definition of the $u$ parameter is different: $u=\sqrt{\tau^2 + \beta^2}$, where
\begin{align}
\tau(t_i) = \frac{t_i-t_0}{\tE}+\delta\tau, \quad \beta(t_i) = u_0 + \delta\beta,
\end{align}
and
\begin{align}
(\delta\tau,\delta\beta) = (\piEvec\cdot\Delta\boldsymbol{s},\piEvec\times\Delta\boldsymbol{s}),
\end{align}
where $\Delta\boldsymbol{s}$ is the projected position of the Sun in the adopted geocentric frame \citep{gould2004}.
For four events from the sample, including the microlens parallax effects improved the fit significantly: OGLE-LMC-09 ($\Delta\chi^2=252.3$), OGLE-LMC-10 ($\Delta\chi^2=129.0$), OGLE-LMC-12 ($\Delta\chi^2=8.2$), and OGLE-LMC-15 ($\Delta\chi^2=11.7$). These objects are likely due to lenses located in the foreground Milky Way disk, as discussed by \citet{mroz2024b} in more detail. The degeneracies in the parallax solutions are discussed in Appendix.

For the events for which good quality $V$-band data were taken during the magnified part of the light curve, we measured the color of the source using the model-independent regression. Source colors are reported in Table~\ref{tab:colors}, and they are generally similar to the colors of events in the baseline, confirming that the majority of observed light originates from the source star. Figure~\ref{fig:cmd} presents a dereddened color--magnitude diagram of one of the observed fields (LMC516.14). Positions of detected events (in the baseline) are indicated by blue circles. Most of the events occurred on giants and subgiants, and a few events are located on the blue end of the main sequence, confirming that the source stars are located in the LMC.

The overall properties of all events (their positions in the sky, timescales, and microlensing parallaxes) are consistent with lenses located in the LMC or in the foreground Milky Way disk. A comparison with predictions from the LMC and Milky Way models is discussed in detail in the companion paper \citep{mroz2024b}.

\subsection{Comparison with Previous Searches for Microlensing Events toward the LMC}

The analyzed dataset, by design, contained the vast majority of OGLE-III light curves previously investigated by \citet{wyrzykowski2011}, who discovered four events. We successfully recovered three of them (OGLE-LMC-04, OGLE-LMC-05, and OGLE-LMC-06). The only remaining event (OGLE-LMC-03) was not selected by our pipeline because a 2000--2009 portion of its light curve was missing in the new database. This event is one of a few examples in which the new OGLE-III image reductions failed (Section~\ref{sec:newdata}). On the other hand, we identified two events (OGLE-LMC-07 and OGLE-LMC-08) that occurred during the OGLE-III phase and were missed by \citet{wyrzykowski2011}.

With the extended OGLE-IV dataset, we also had an opportunity to verify the candidate microlensing events detected by previous searches. If objects underwent additional outbursts or exhibited some other variability, their microlensing interpretation is unlikely to be correct. \citet{wyrzykowski2009} discovered two candidate events in the OGLE-II data from 1996 to 2000. The updated light curve of OGLE-LMC-01 showed no variability in data collected after 2001, but OGLE-LMC-02 exhibited three additional outbursts: in 2010 March, 2014 December, and 2019 February. The latter object shares several similarities to known transient supersoft X-ray sources in the Magellanic Clouds \citep{maccarone2019,mroz2023} and will be discussed in more detail elsewhere. Because OGLE-LMC-02 is not a microlensing event, the estimates of the optical depth and event rate from OGLE-II data by \citet{wyrzykowski2009} need to be revised: $\tau = (0.28 ^{+0.64}_{-0.23}) \times 10^{-7}$ and $\Gamma = (1.1 ^{+2.5}_{-0.9}) \times 10^{-7}\ \mathrm{yr}^{-1}$.

One of the candidate events found by \citet{tisserand2007} in the full sample of EROS stars (EROS2-SMC-2) underwent a low-amplitude outburst in 2017 November. The EROS limits on the microlensing optical depth remain unchanged because \citet{tisserand2007} did not use this event to determine the optical depth.

Among 17 candidate events announced by \citet{alcock2000b} using the MACHO data, two objects showed outbursts in the OGLE-III and OGLE-IV data: MACHO-LMC-7 (in 2005 March, 2007 January, 2009 February, 2010 November, 2012 January, 2015 November, 2018 September) and MACHO-LMC-23 (in 2001 November). The variability of MACHO-LMC-23 was already noticed by \citet{bennett2005b}, \citet{tisserand2007}, and \citet{wyrzykowski2011}. \citet{wyrzykowski2011} flagged MACHO-LMC-7 as a variable object. Additionally, we found that three candidate events (MACHO-LMC-08, MACHO-LMC-18, and MACHO-LMC-27) show periodic variability in the baseline (with periods $2.310617 \pm 0.000010$, $7.02772 \pm 0.00015$, and $1.532534 \pm 0.000025$ days, respectively), indicating that outbursts detected by MACHO were likely of stellar origin. We found that the centroids of periodic variable stars measured on subtracted images from the DIA pipeline matched precise coordinates of MACHO-LMC-08 and MACHO-LMC-18 determined by \citet{nelson2009}. Although some microlensing events with periodic variability in the baseline are known toward the Galactic bulge \citep[e.g.,][]{wyrzykowski2006}, they are 2 orders less common than microlensing events with a constant baseline. The probability of observing three variable-baseline events given the expected 0.2 events in a sample of 17 events is only $10^{-3}$. All five objects are likely to be false positives, demonstrating a high contamination rate ($\approx 30\%$) of the MACHO sample. Removing these objects from the sample used to derive the microlensing optical depth by \citet{alcock2000b} lowers it by almost 40\%, to $0.74 \times 10^{-7}$. We note that all five events identified as variable stars in this work were considered to be ``unconfirmed'' or ``rejected'' in follow-up studies by \citet{bennett2005b} and \citet{bennett2005}.

Six out of 16 analyzed microlensing events (OGLE-LMC-07, OGLE-LMC-09, OGLE-LMC-10, OGLE-LMC-13, OGLE-LMC-14, OGLE-LMC-18) were observed in 1996--2000 by the OGLE-II survey. Light curves of all six events were flat. Finally, we retrieved the archival MACHO photometry for all 16 microlensing events analyzed in this work. The MACHO data spanned the period from 1992 July to 1999 December, so they are complementary to the OGLE-III and OGLE-IV data sets and extend the light-curve coverage to nearly 28~yr. We found no outbursts or brightenings in the archival MACHO light curves, consistent with the classification of all analyzed objects as microlensing events.

\section{Number of Source Stars}
\label{sec:stars}

The number of stars detected on OGLE reference images is not equivalent to the number of microlensing source stars that enter the microlensing optical depth calculation (Equation~(\ref{eq:tau})). First, in ground-based, seeing-limited images, two or more stars may be blended and cannot be resolved. Second, some stars observed by OGLE are in the foreground Milky Way disk, and the probability that they would be lensed by compact objects in the Milky Way halo is negligible. Therefore, they should not enter the optical depth calculation.

To quantify the effects of blending on star counts, we used the archive of high-quality stellar photometry from the Hubble Space Telescope (HST) by \citet{holtzman2006}. The HST images were deeper and had a superb resolution compared to ground-based images, so the vast majority of stars were resolved. The HST images in the $F814W$ filter (closely matching the standard $I$-band filter) were available for 64 fields, spanning over a factor of 20 in stellar surface density, from low-density regions in the outer part of the LMC, to the densest fields in the LMC bar.

We calculated the surface density of stars per square arcmin on the HST images $\Sigma^{HST}$ and compared it to the surface density of stars per square arcmin detected on OGLE reference images $\Sigma^{\rm OGLE}$. Following \citet{wyrzykowski2011}, we called the ratio ${\Sigma^{HST}}/{\Sigma^{\rm OGLE}}$ a correction factor (CF), because it allowed us to convert the raw OGLE star counts to the actual number of stars that may be microlensed. We calculated stellar surface density separately for three magnitude ranges: for stars brighter than $I=21$, $I=21.5$, and $I=22$ mag.

As an example, Figure~\ref{fig:corr} shows the CF as a function of the OGLE star density for stars brighter than $I=21$ mag. We found that it was related to the observed surface density of stars by a formula:
\begin{equation}
\mathrm{CF} = \frac{\Sigma^{HST}}{\Sigma^{\rm OGLE}} = c + \exp\left(a(\log\Sigma^{\rm OGLE}-2.5)+b\right),
\label{eq:cf}
\end{equation}
where $a=4.04 \pm 0.21$, $b=-2.246 \pm 0.072$, and $c=1.00 \pm 0.02$ are the best-fit coefficients.

We derived similar fits for stars brighter than $I=21.5$ mag, $a=3.985 \pm 0.077$, $b=-1.527 \pm 0.027$, and $c=1.03 \pm 0.02$; and for stars brighter than $I=22$ mag, $a=2.746 \pm 0.032$, $b=-0.489 \pm 0.010$, and $c=1.05 \pm 0.01$. These relations allowed us to convert the raw star counts into the total number of possible microlensing source stars. By comparing the predicted number of stars with the actual HST observations, the accuracy of the determination of star counts was estimated to be 3\% (for stars brighter than $I=21$), 7\% (for stars brighter than $I=21.5$), and 11\% (for stars brighter than $I=22$).

In the next step, we used the star counts in fields located in the outer parts of the LMC region, at $\mathrm{R.A.} \gtrsim 6^{\rm h}30^{\rm m}$ or $\mathrm{Decl.} \gtrsim -60^{\circ}$, to estimate the Milky Way foreground. We found that the surface density of foreground stars depends on the Galactic latitude $b$ as follows:
\begin{equation}
\log \Sigma^{\rm foreground} = \alpha + \beta (|b| - 30^{\circ}),
\label{eq:foreground}
\end{equation}
where $\alpha$ and $\beta$ are the best-fit coefficients. For stars brighter than $I=21$ mag, $\alpha=0.846 \pm 0.016$, $\beta=-0.0231 \pm 0.0019$; for stars brighter than $I=21.5$ mag, $\alpha=0.873 \pm 0.018$, $\beta=-0.0231 \pm 0.0022$; for stars brighter than $I=22$ mag, $\alpha=0.893 \pm 0.019$, $\beta = -0.0230 \pm 0.0023$.

The final number of source stars observed in a given field was calculated by multiplying the number of stars detected on OGLE reference images by the CF (depending on stellar density, Equation~(\ref{eq:cf})), and subtracting the foreground stars (using Equation~(\ref{eq:foreground})). The results are presented in Table~\ref{tab:stars} in which we report $N_{\rm total}$ -- the total number of stars (after correcting the raw star counts for blending), and $N_{\rm corr}$ -- the number of possible microlensing source stars after removing foreground Milky Way sources. Figure~\ref{fig:stars} shows the surface density of source stars brighter than $I=22$ mag.

\section{Detection Efficiency}
\label{sec:eff}

We ran extensive simulations to measure the detection efficiency of microlensing events as a function of their Einstein timescale in our experiment. Our method was similar to that described by \citet{mroz2019b}. We created synthetic light curves of microlensing events by injecting the simulated signal into light curves from the database. The simulated light curves, therefore, had similar noise properties, the same sampling, and the same outliers as the original light curves. We subsequently ran the event and detection modeling pipeline (Section~\ref{sec:events}) on the synthetic light curves and calculated the fraction of events that pass all selection criteria (Table~\ref{tab:cuts}).

The time of the peak $t_0$ was drawn from a uniform distribution from the range $2,452,000 \leq t_0 \leq 2,459,000$ (from 2001 April to 2020 May) for fields observed during the OGLE-III and OGLE-IV phases, and $2,455,000 \leq t_0 \leq 2,459,000$ (from 2009 June to 2020 May) for fields observed during OGLE-IV only. The Einstein timescale $\tE$ was drawn from a log-uniform distribution from the range 1 to $10^4$ days, and the impact parameter $u_0$ was drawn from a uniform distribution from the range $0 \leq u_0 \leq 1$. The blending parameter $f_{\rm s}$, depending on the mean magnitude of the star, was drawn from the empirical distribution derived by cross-matching OGLE and HST stars (Section~\ref{sec:stars}). To take into account binary-lens events, which our automated searches may miss, we lowered the detection efficiency by 10\% \citep{wyrzykowski2011,mroz2019b}.

We simulated 30,000 events per detector, almost $10^6$ events per field, and calculated the detection efficiency in 25 identical bins in $\log\tE$. The example detection efficiency curves (for field LMC501) are shown in Figure~\ref{fig:eff}. For fields observed during OGLE-III and OGLE-IV phases, the sensitivity peaks at $\tE = 2.5$\,yr and declines by 50\% at $\tE = 13.8$\,yr. For fields observed during OGLE-IV only, the sensitivity peak is at $\tE = 1.2$\,yr, and the detection efficiency falls by 50\% at $\tE=6.2$\,yr. We calculated detection efficiencies averaged for sources brighter than $I=21$, $I=21.5$, and $I=22$ mag.

\section{Optical Depth}
\label{sec:tau}

The microlensing optical depth $\tau$ and event rate $\Gamma$ were evaluated using the formulae
\begin{align}
\begin{split}
\tau &= \frac{\pi}{2N_{\rm s}}\sum_i\frac{1}{\Delta T_i}\frac{t_{{\rm E}, i}}{\varepsilon(t_{{\rm E}, i})},\\
\Gamma &= \frac{1}{N_{\rm s}}\sum_i\frac{1}{\Delta T_i}\frac{1}{\varepsilon(t_{{\rm E}, i})},
\end{split}
\end{align}
where $\Delta T=7000$ days, or $\Delta T = 4000$ days if the event was observed during both OGLE-III and OGLE-IV phases or during OGLE-IV only, respectively. $N_{\rm s}=78.7 \times 10^6$ is the number of source stars in the LMC monitored in our experiment (Table~\ref{tab:stars}). To take into account uncertainties on the event timescale, we replaced the terms $t_{{\rm E}, i}/\varepsilon(t_{{\rm E}, i})$ and $1/\varepsilon(t_{{\rm E}, i})$ by the mean over $N_i$ samples from the posterior distribution of $t_{{\rm E},i}$:
\begin{align}
\begin{split}
\frac{t_{{\rm E}, i}}{\varepsilon(t_{{\rm E}, i})} &\longrightarrow \frac{1}{N_i}\sum_{k=1}^{N_i}\frac{t_{{\rm E}, ik}}{\varepsilon(t_{{\rm E}, ik})},\\
\frac{1}{\varepsilon(t_{{\rm E},i})} &\longrightarrow \frac{1}{N_i}\sum_{k=1}^{N_i}\frac{1}{\varepsilon(t_{{\rm E},ik})}.
\end{split}
\end{align}
We used only events passing our automated selection criteria to calculate the optical depth and event rate. Our sample includes 13 such events. The total optical depth toward the OGLE-IV fields is estimated at $\tau = (0.121 \pm 0.037) \times 10^{-7}$ and the total event rate is $\Gamma = (0.74 \pm 0.25) \times 10^{-7}\,\mathrm{yr}^{-1}\,\mathrm{star}^{-1}$. The error bars were calculated using the method described by \citet{han1995_stat}. There is an additional systematic error of 11\% related to the accuracy of finding the number of source stars (Section~\ref{sec:stars}). The contribution of each event to the total optical depth and event rate is reported in Table~\ref{tab:tau}. 

If we had used the median value from the posterior distribution of $\tE$ (and the detection efficiency evaluated at the median timescale), instead of using the full posterior, then the total optical depth was estimated at $\tau = (0.112 \pm 0.035) \times 10^{-7}$ and the event rate was $\Gamma = (0.69 \pm 0.23) \times 10^{-7}\,\mathrm{yr}^{-1}\,\mathrm{star}^{-1}$. These values are about 7\% lower than those calculated using the full posterior distribution; we consider them as less accurate.

\section{Discussion}
\label{sec:discussion}

This work represents a significant improvement over the previous efforts to measure the optical depth and event rate toward the LMC. The previous determinations of these quantities by EROS \citep{tisserand2007}, OGLE-II \citep{wyrzykowski2009}, and OGLE-III surveys \citep{wyrzykowski2011} were based on small number statistics (either one or two events). Combining the OGLE-III and OGLE-IV datasets enabled us to create homogeneous light curves covering nearly 20~yr of photometric monitoring. The analysis of this vast dataset, containing the light curves of 62 million stars, led to the discovery of 16 microlensing events. Thirteen passed our strict selection criteria, allowing us to measure the microlensing optical depth and event rate toward the LMC with unprecedented precision.

Our measurements are in good agreement with the results reported by \citet{wyrzykowski2011} from the OGLE-III survey, who found $\tau = (0.16 ^{+0.21}_{-0.10}) \times 10^{-7}$. Although they did not explicitly report the event rate measurement, the data presented by \citet{wyrzykowski2011} enabled us to find $\Gamma = (1.14 ^{+1.50}_{-0.74}) \times 10^{-7}\,\mathrm{yr}^{-1}\,\mathrm{star}^{-1}$. These measurements are formally consistent within the reported error bars (which are large and asymmetric). However, the best estimates are 30\%--50\% higher than reported here (Section~\ref{sec:tau}). The difference in sky coverage between the OGLE-III and OGLE-IV fields can explain this difference. The optical depth and event rate were slightly higher because the OGLE-III fields covered a smaller sky area than OGLE-IV, centered on the inner LMC. When we restricted our sample to events located within the OGLE-III fields, then the optical depth and event rate were $\tau = (0.150 \pm 0.050) \times 10^{-7}$ and $\Gamma = (0.83 \pm 0.32) \times 10^{-7}\,\mathrm{yr}^{-1}\,\mathrm{star}^{-1}$, reaching a better agreement with the \citet{wyrzykowski2011} findings.

The findings presented in this work also confirm the results from EROS, OGLE-II, and OGLE-III that the optical depth toward the LMC is about an order of magnitude smaller than that reported by the MACHO collaboration \citep{alcock2000b,bennett2005}. \citet{besla2013} argued that this discrepancy results from different samples of stars analyzed in each experiment: EROS and OGLE-III were concentrated on bright stars ($R<19.7$ for EROS, $I \leq 18.8$ for OGLE-III), whereas MACHO analyzed stars as faint as $V=22$. \citet{besla2013} proposed that such faint sources may originate from a population of tidally stripped stars from the Small Magellanic Cloud located behind the LMC.
In this work, we studied a sample of all stars observed by OGLE-IV and confirmed the low value of the optical depth. Moreover, we found that at least five candidate events reported by MACHO (30\% of their sample) are likely due to stellar variability rather than microlensing. If the contamination rate were higher than might be inferred from these five spurious events, then this might account for additional parts of the discrepancy. However, at present, we have no evidence for such higher contamination.

Our primary motivation for carrying out this work and combining the OGLE-III and OGLE-IV datasets was the fact that all previous microlensing experiments were not sensitive to events with timescales longer than $\tE \gtrsim 2.5\mbox{--}3$ yr, raising the possibility that such long-timescale events were not accounted for in the calculation of the microlensing optical depth. We did not find any such long-timescale events in our data, confirming that the optical depth toward the LMC is significantly smaller than that expected from a dark matter halo in the form of compact objects. Our conclusions about the abundance of compact objects in dark matter are presented in detail in a companion paper \citep{mroz2024b}. In short, we found that all detected events can be explained by the lensing by stars in the foreground Milky Way or in the LMC itself, without the need to invoke dark matter in the form of compact objects. We found that compact objects in the mass range from $1.8\times 10^{-4}\,M_{\odot}$ to $6.3\,M_{\odot}$ cannot compose more than 1\% of dark matter, and objects in the mass range from $1.3\times 10^{-5}\,M_{\odot}$ to $860\,M_{\odot}$ cannot make up more than 10\% of dark matter.

An exact origin of lenses can be traced with follow-up observations. Revisiting the events with adaptive-optics observations may reveal the source and the lens separated in the sky. Combining the lens flux measurements with additional information, for example, from the microlens parallax effects, should enable us to pinpoint the mass and distance to the lens.

The data presented in this paper are publicly available at \url{https://www.astrouw.edu.pl/ogle/ogle4/LMC_OPTICAL_DEPTH} and Zenodo \citep{mroz_zenodo}.

\section*{Acknowledgements}

We thank the referee, Prof.~Andrew Gould, for very constructive comments that helped us to improve the presentation of our results. We thank all the OGLE observers for their contribution to the collection of the photometric data over the decades. This research was funded in part by National Science Centre, Poland, grant OPUS 2021/41/B/ST9/00252 awarded to P.M.

\bibliographystyle{aasjournal}
\bibliography{pap}

\appendix
\section{Degeneracies in parallax models}

We carried out a grid search in the $(\piEN,\piEE)$ plane to study microlensing parallax degeneracies in the detected events. The best-fit models in each grid point were found by minimizing the $\chi^2$ function (Equation~(\ref{eq:chi2})) with the prior on negative blending (Equation~(\ref{eq:prior})) using the MCMC code by \citet{foreman2013}. We kept $(\piEN,\piEE)$ fixed, but other parameters were allowed to vary. The grid consisted of $251 \times 251$ points uniformly spread over the range $(-5 \leq \piEN \leq 5, -5 \leq \piEE \leq 5)$.

The likelihood contours in the $(\piEN,\piEE)$ plane are shown in Figure~\ref{fig:parallaxes} for six events (OGLE-LMC-09, OGLE-LMC-10, OGLE-LMC-11, OGLE-LMC-12, OGLE-LMC-15, and OGLE-LMC-16), for which meaningful contraints on $\piEvec$ could be obtained. The red, orange, yellow, lime, green, cyan, blue, and dark blue colors mark $\Delta\chi^2=1,4,9,15,25,36,49,64$, respectively. The limits on the microlensing parallax are weak for the remaining events from our sample (which have relatively short timescales and/or are faint).

We separately plotted solutions with positive and negative values of $u_0$, which are known to be almost perfectly symmetric \citep[e.g.,][]{smith2003}. We also plotted the directions of $(\pi_{\rm E,\parallel},\pi_{\rm E,\perp})$, which are parallel and perpendicular to the apparent acceleration of the Sun (projected on the sky) in an Earth frame, as defined by \citet{gould2004}.

The main degeneracies in the analyzed light curves may be explained by the jerk-parallax degeneracy found by \citet{gould2004}. \citet{gould2004} demonstrated that, in the limit $|u_0| \ll 1$, if $\piEvec$ is one solution, then $\boldsymbol{\pi}_{\rm E}'$ with
\begin{equation}
(\pi'_{\rm E,\parallel}, \pi'_{\rm E,\perp}) = (\pi_{\rm E,\parallel},-(\pi_{\rm E,\perp}+\pi_{j,\perp}))
\end{equation}
is also a solution. Here, the ``jerk parallax'' $\boldsymbol{\pi}_j$ is defined by
\begin{equation}
\boldsymbol{\pi}_j = \frac{4}{3}\frac{\boldsymbol{j}}{\alpha^2\tE},
\end{equation}
where $\boldsymbol{\alpha}$ and $\boldsymbol{j}$ are the apparent acceleration and jerk of the Sun (projected on the sky) relative to the Earth, each divided by an astronomical unit.

Figure~\ref{fig:parallaxes} shows that the degenerate parallax solutions are aligned along the axis perpendicular to the projected acceleration of the Sun. The microlensing parallax vector components parallel to the projected acceleration of the Sun ($\pi_{\rm E,\parallel}$) are similar for all possible degenerate solutions for a given event. These characteristics are qualitatively consistent with \citet{gould2004} predictions for the jerk-parallax degeneracy.

To provide a more quantitative picture, we reported the coordinates of the local $\chi^2$ grid minima, $(\pi_{\rm E,\parallel}, \pi_{\rm E,\perp})$ and $(\pi'_{\rm E,\parallel}, \pi'_{\rm E,\perp})$, in Table~\ref{tab:paral} (for $u_0>0$ solutions). In all cases, $\pi_{\rm E,\parallel} \approx \pi'_{\rm E,\parallel}$ and $\pi_{\rm E,\perp} + \pi'_{\rm E,\perp} \approx - \pi_{j,\perp}$, as expected for the jerk-parallax degeneracy. The degeneracy seems to be broken for OGLE-LMC-09 ($\Delta\chi^2 = 11.8$), OGLE-LMC-10 ($\Delta\chi^2=16.7$), OGLE-LMC-12 ($\Delta\chi^2=13.0$), and OGLE-LMC-15 ($\Delta\chi^2=38.5$), for which only one of the solutions is preferred.

We also note the presence of narrow local $\chi^2$ minima for OGLE-LMC-11 and OGLE-LMC-16 (at $(\piEN,\piEE) = (0.52,0.36)$ and $(-0.08,-1.00)$, respectively). These solutions are characterized by small values of the impact parameter ($|u_0| \approx 0$) and large blending ($f_{\rm s} \approx 0$), and are likely caused by a previously unrecognized multiparameter ($u_0,\tE,f_{\rm s},\piEvec$) degeneracy.

\newpage
\clearpage

\begin{figure*}
\center
\includegraphics[width=0.49\textwidth]{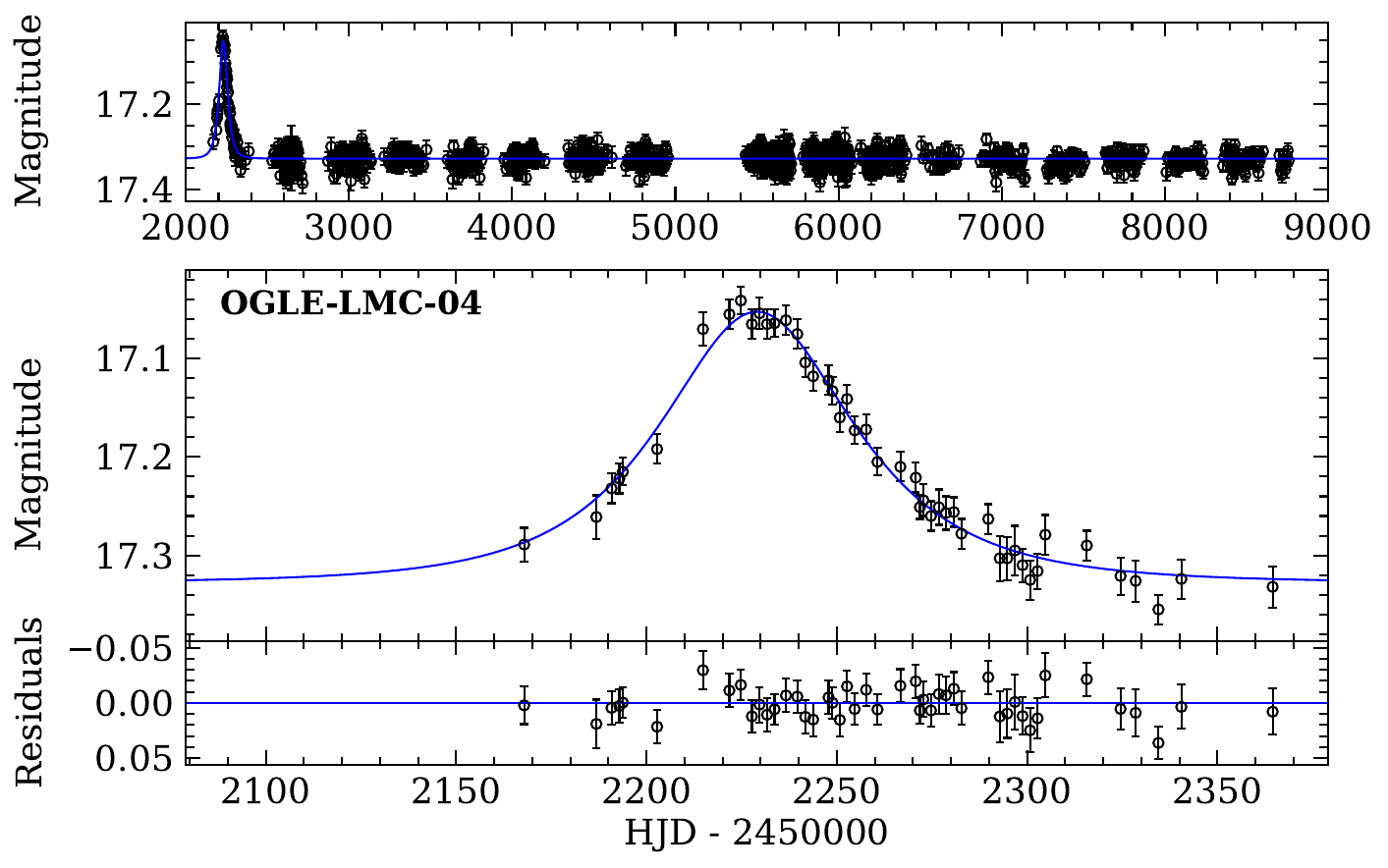}
\includegraphics[width=0.49\textwidth]{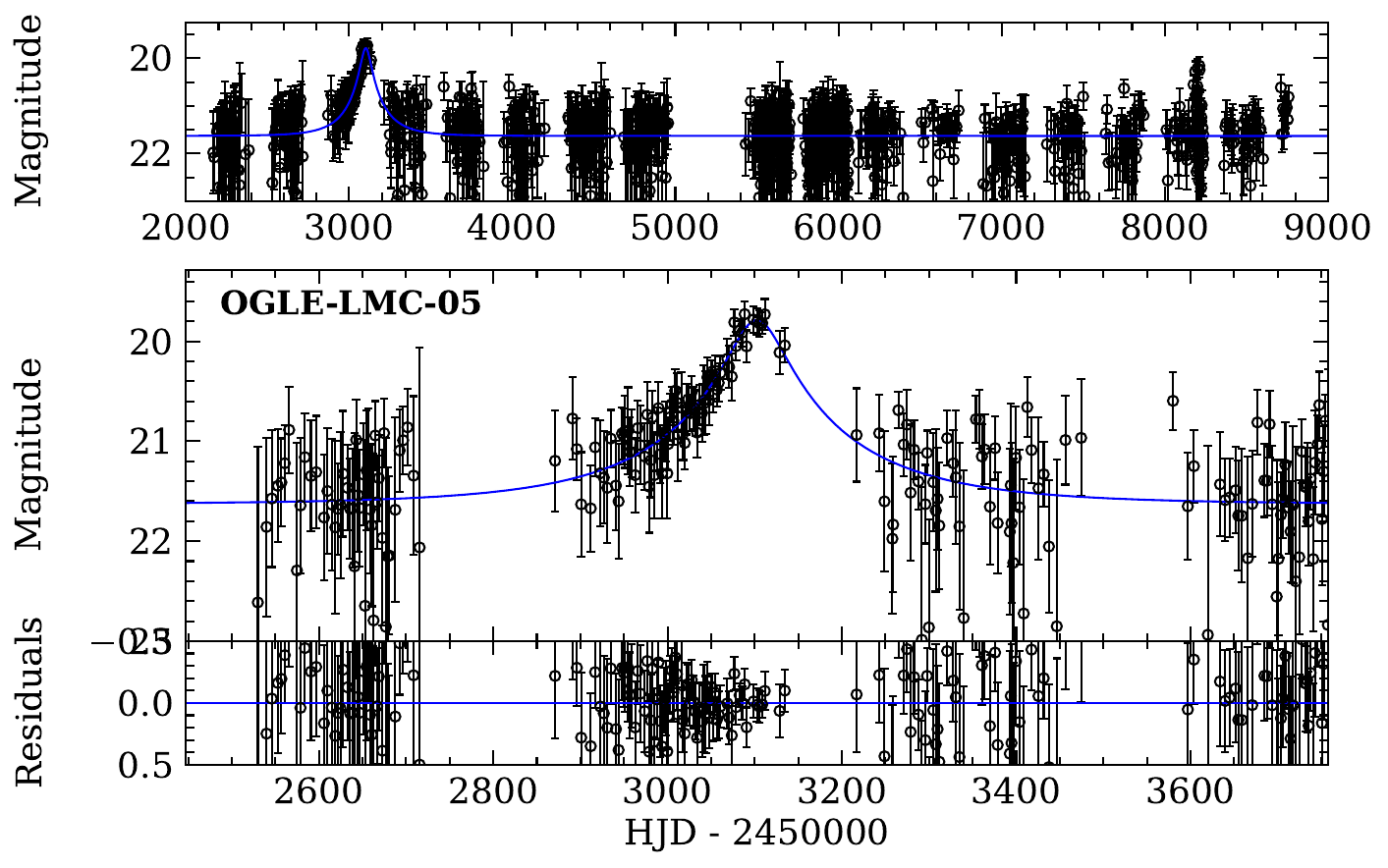}
\includegraphics[width=0.49\textwidth]{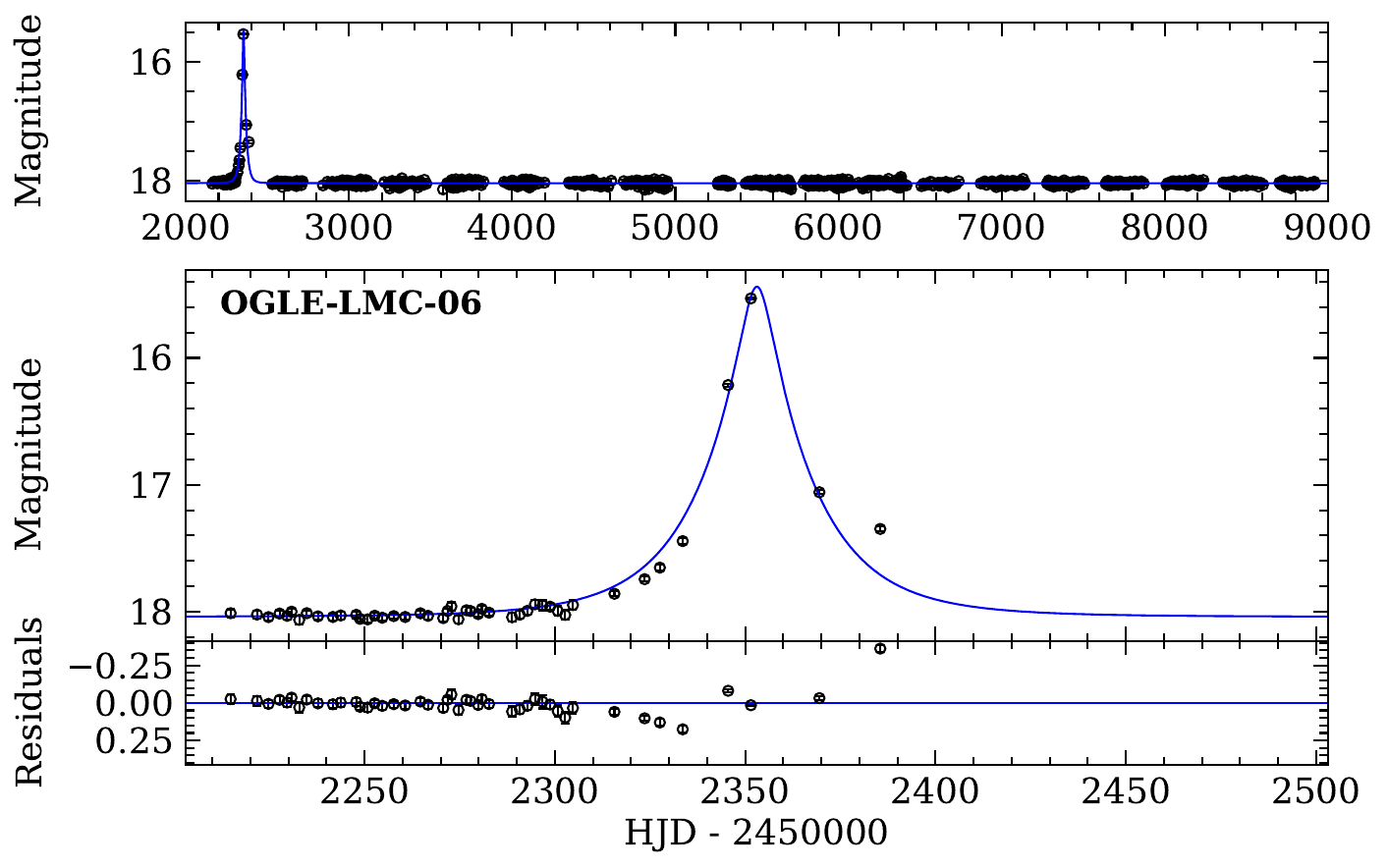}
\includegraphics[width=0.49\textwidth]{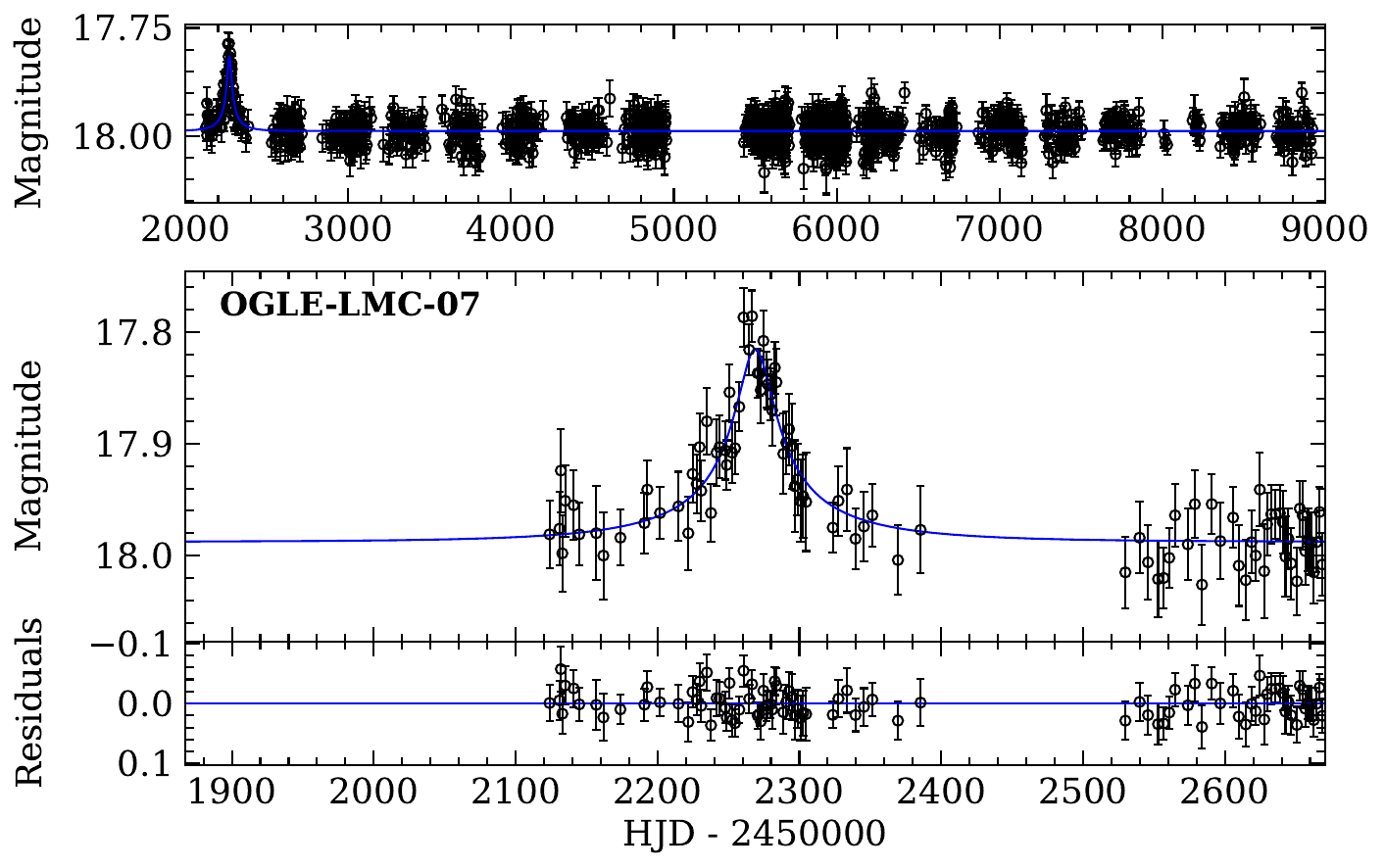}
\includegraphics[width=0.49\textwidth]{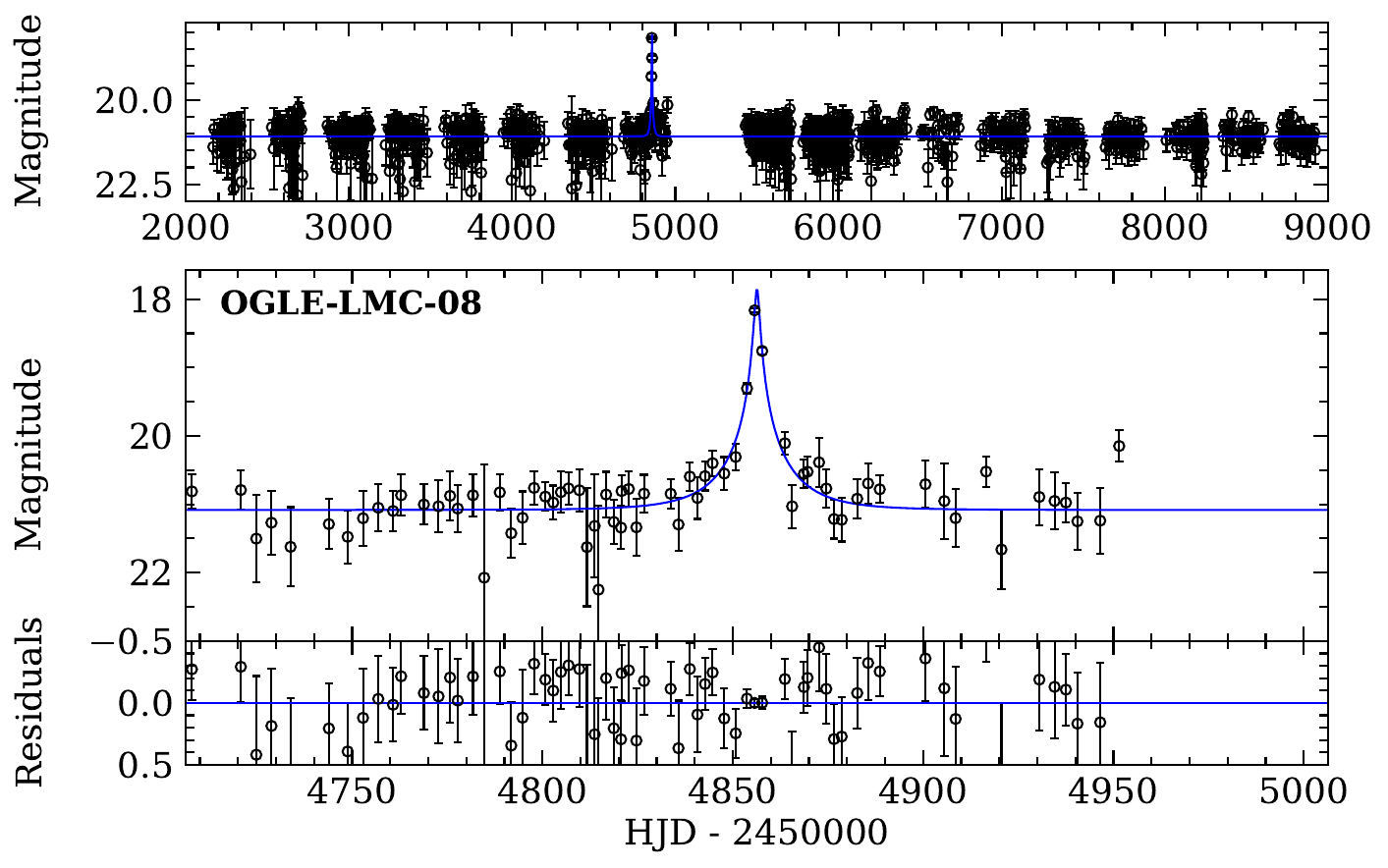}
\includegraphics[width=0.49\textwidth]{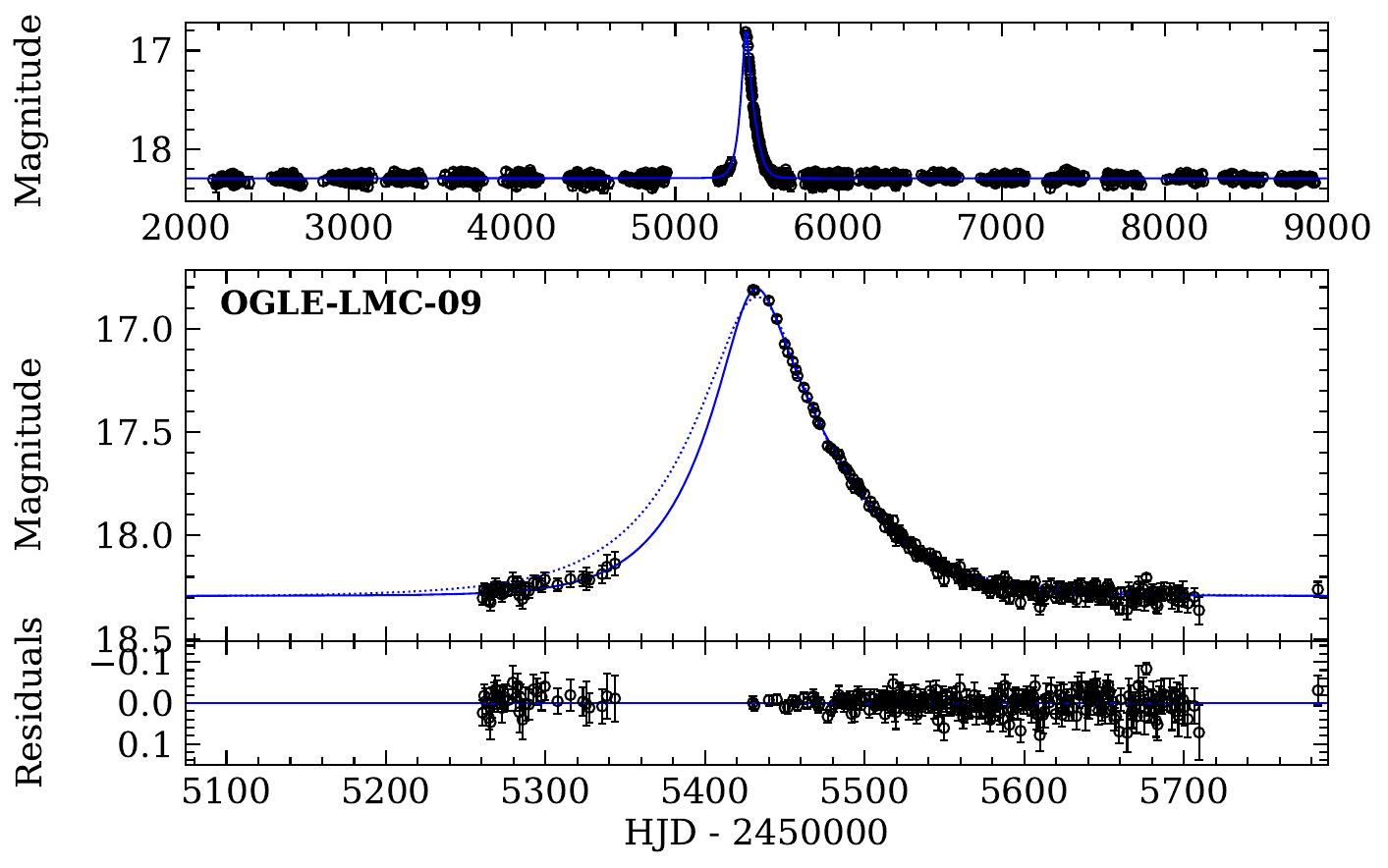}
\includegraphics[width=0.49\textwidth]{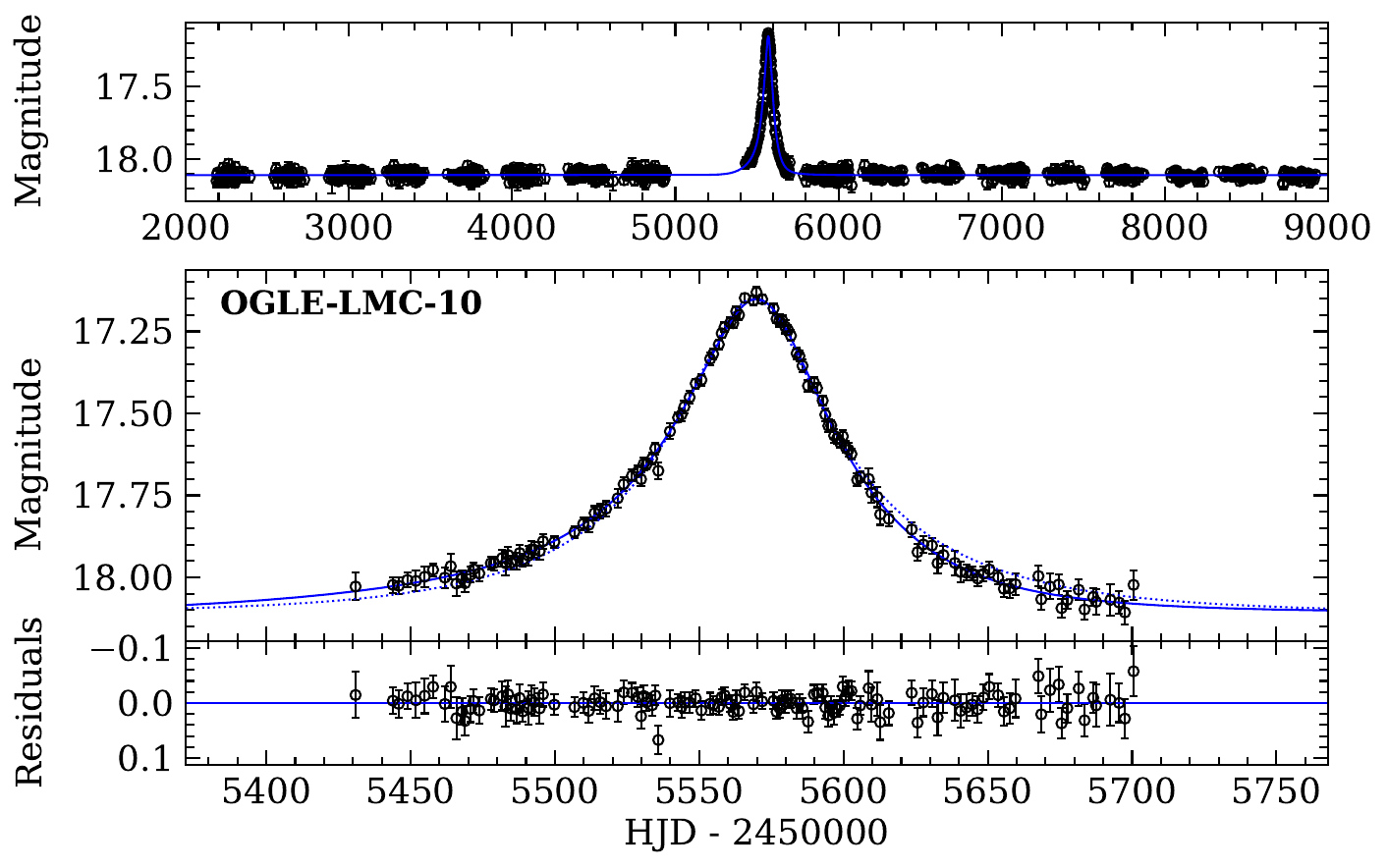}
\includegraphics[width=0.49\textwidth]{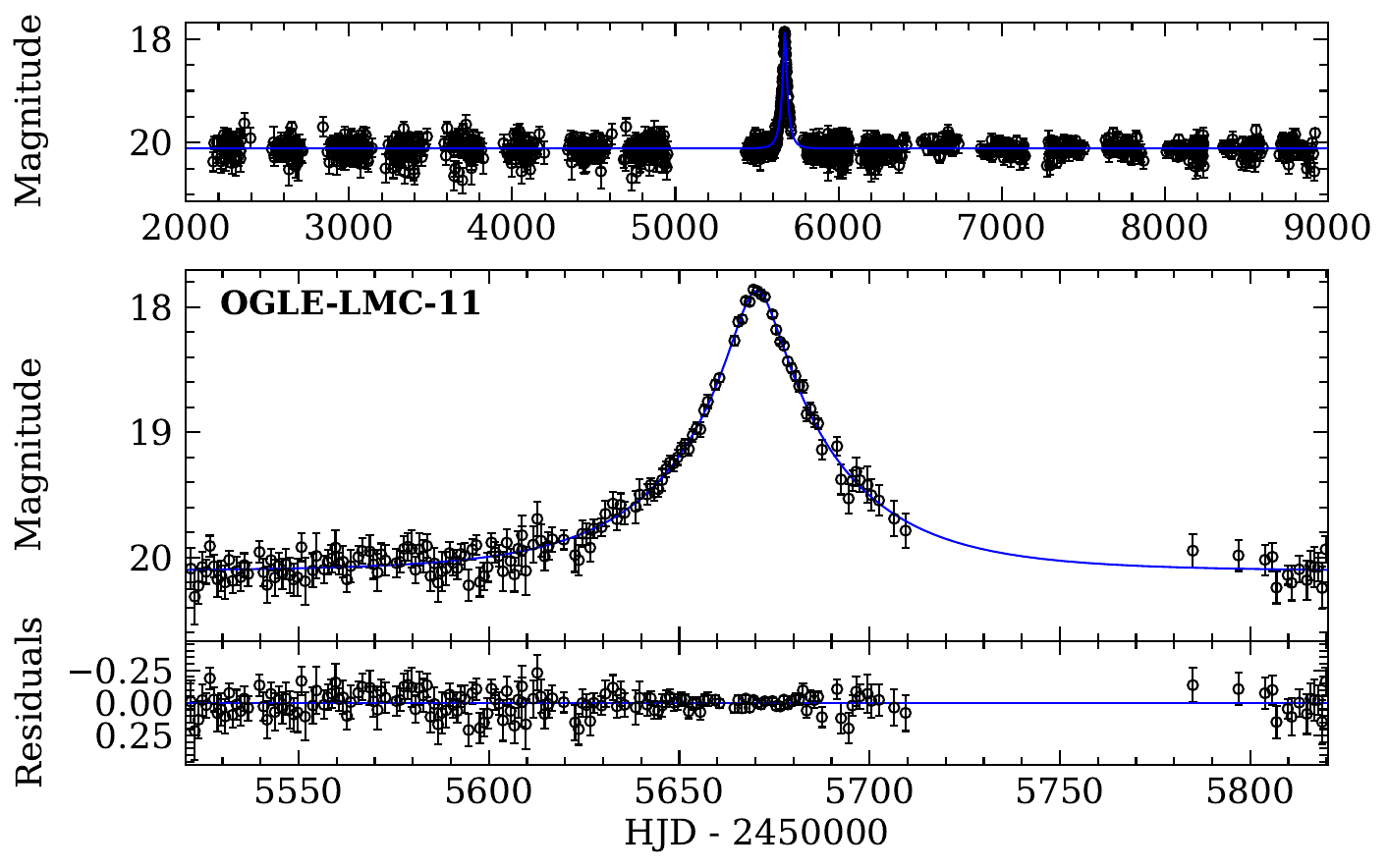}
\caption{Light curves of detected events. The solid blue line shows the best-fit PSPL model (with or without the parallax). For events with detected annual parallax effects, the dotted line shows the best-fit model without the parallax.}
\label{fig:light_curves}
\end{figure*}

\addtocounter{figure}{-1}

\begin{figure*}
\center
\includegraphics[width=0.49\textwidth]{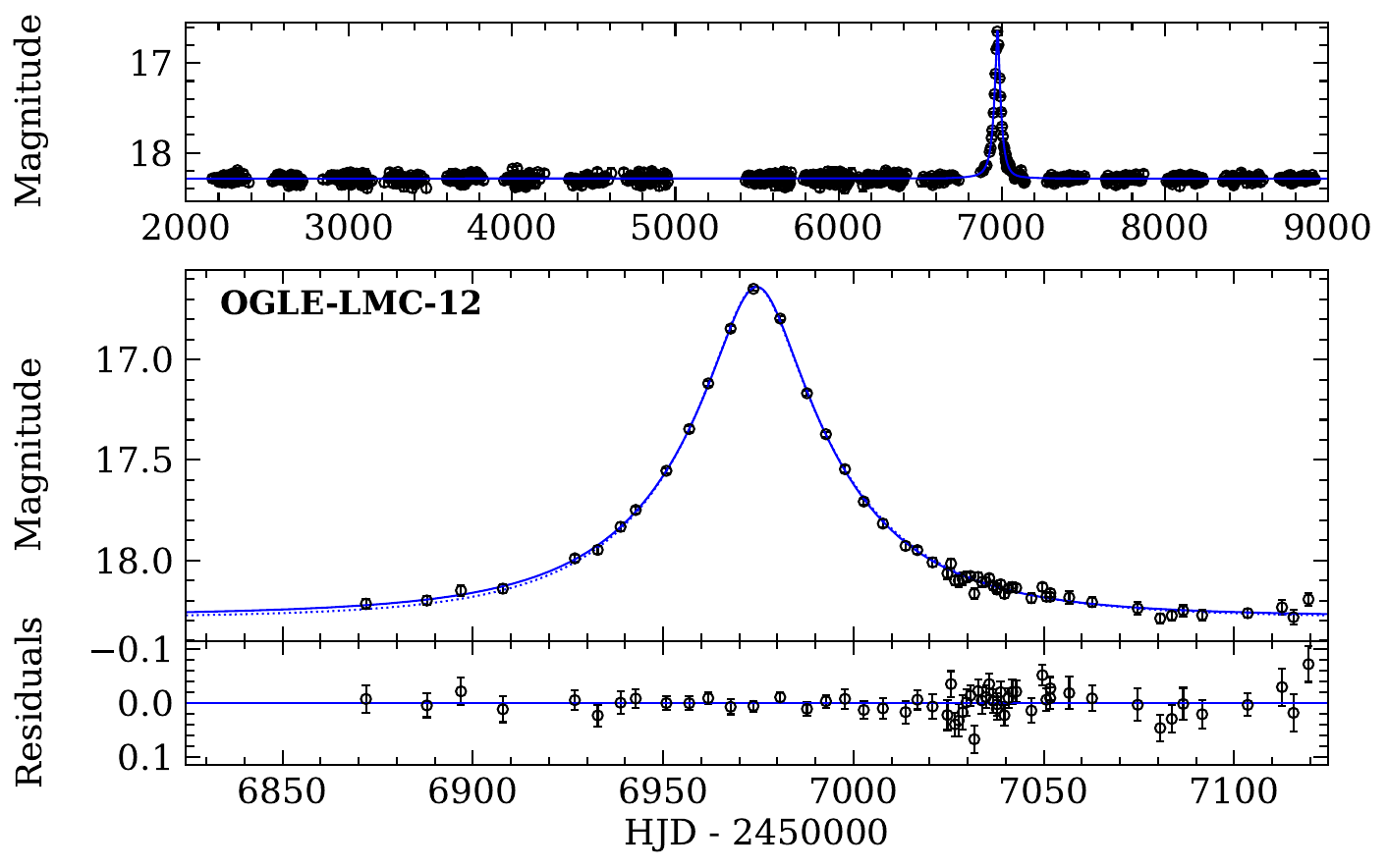}
\includegraphics[width=0.49\textwidth]{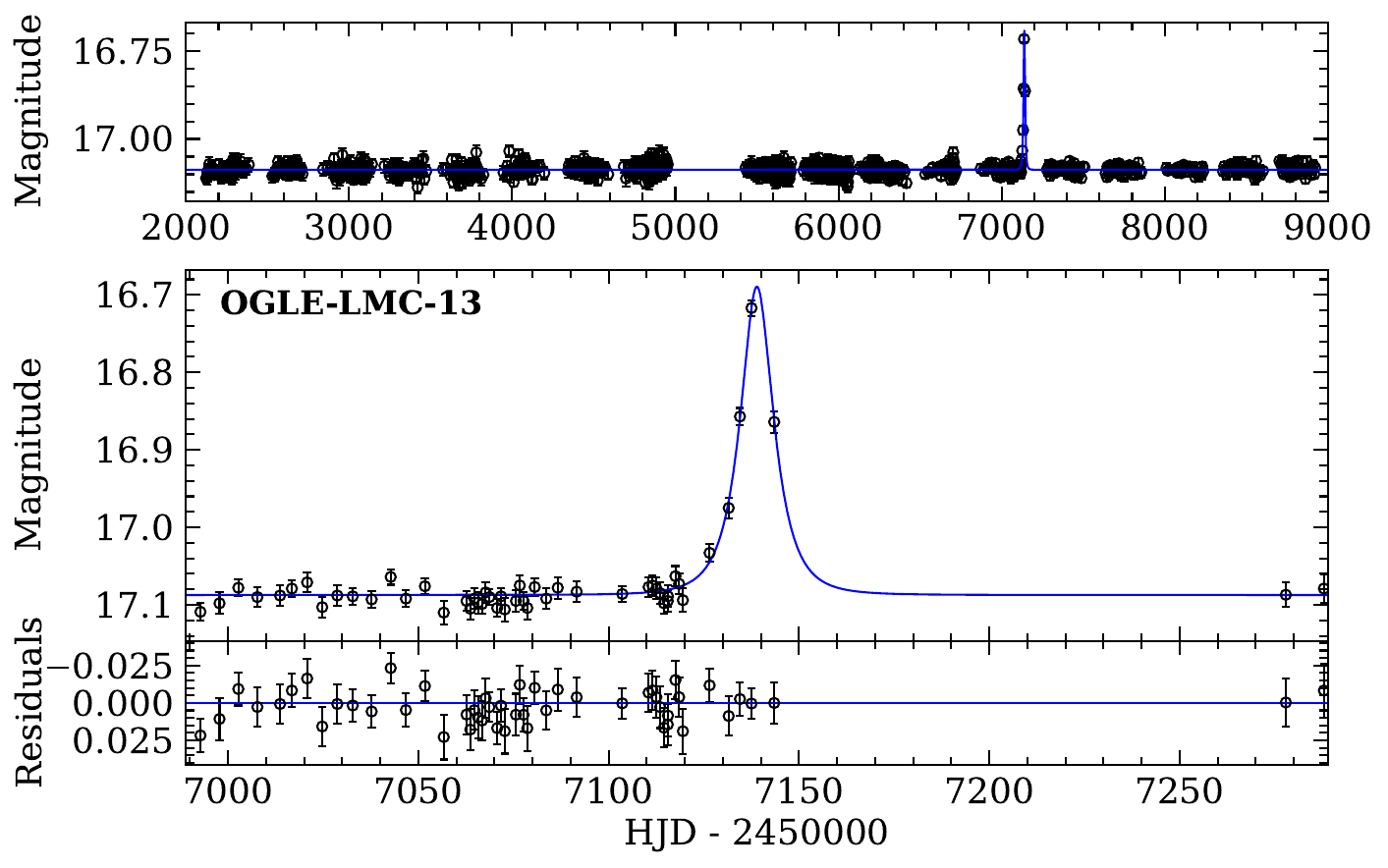}
\includegraphics[width=0.49\textwidth]{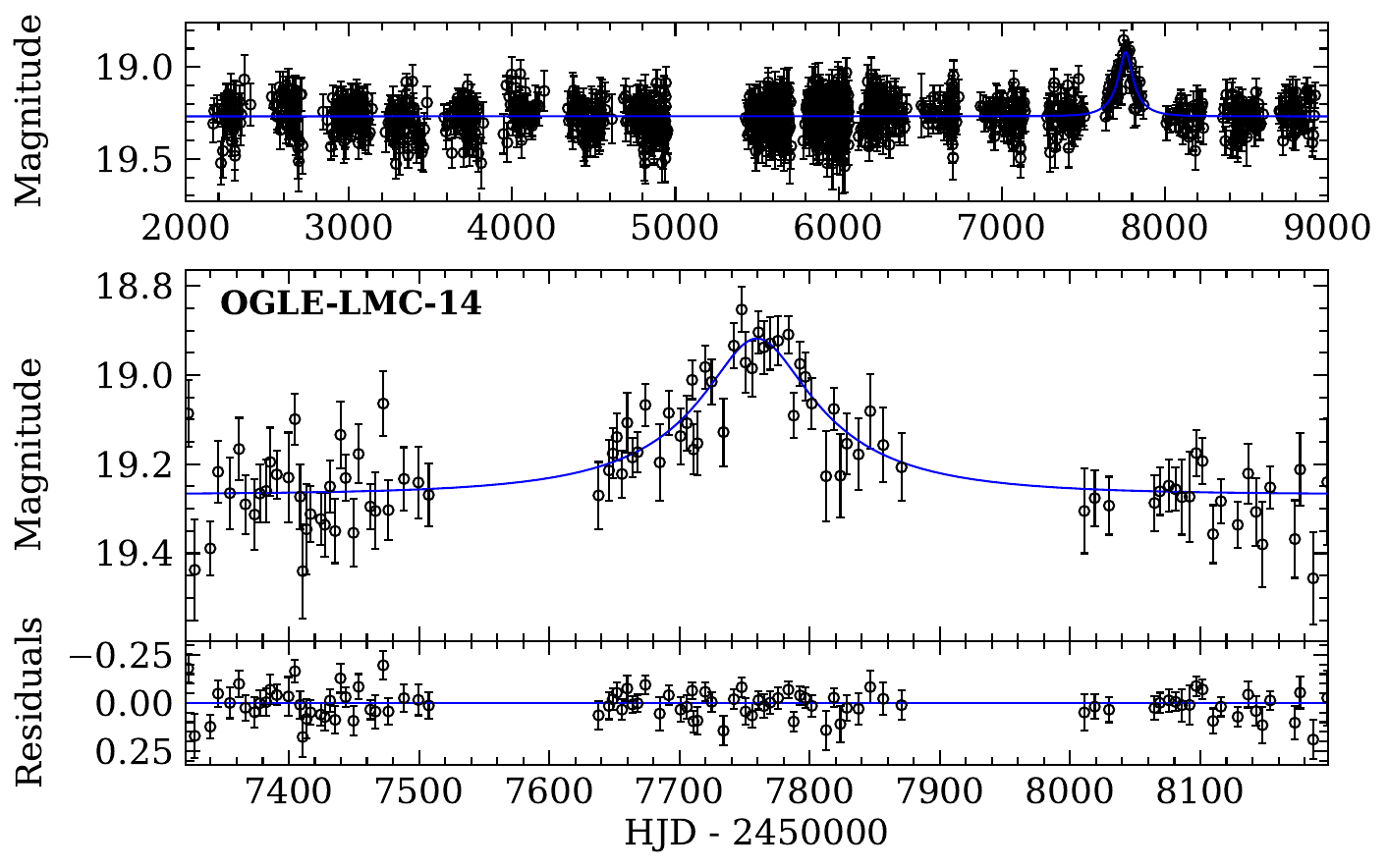}
\includegraphics[width=0.49\textwidth]{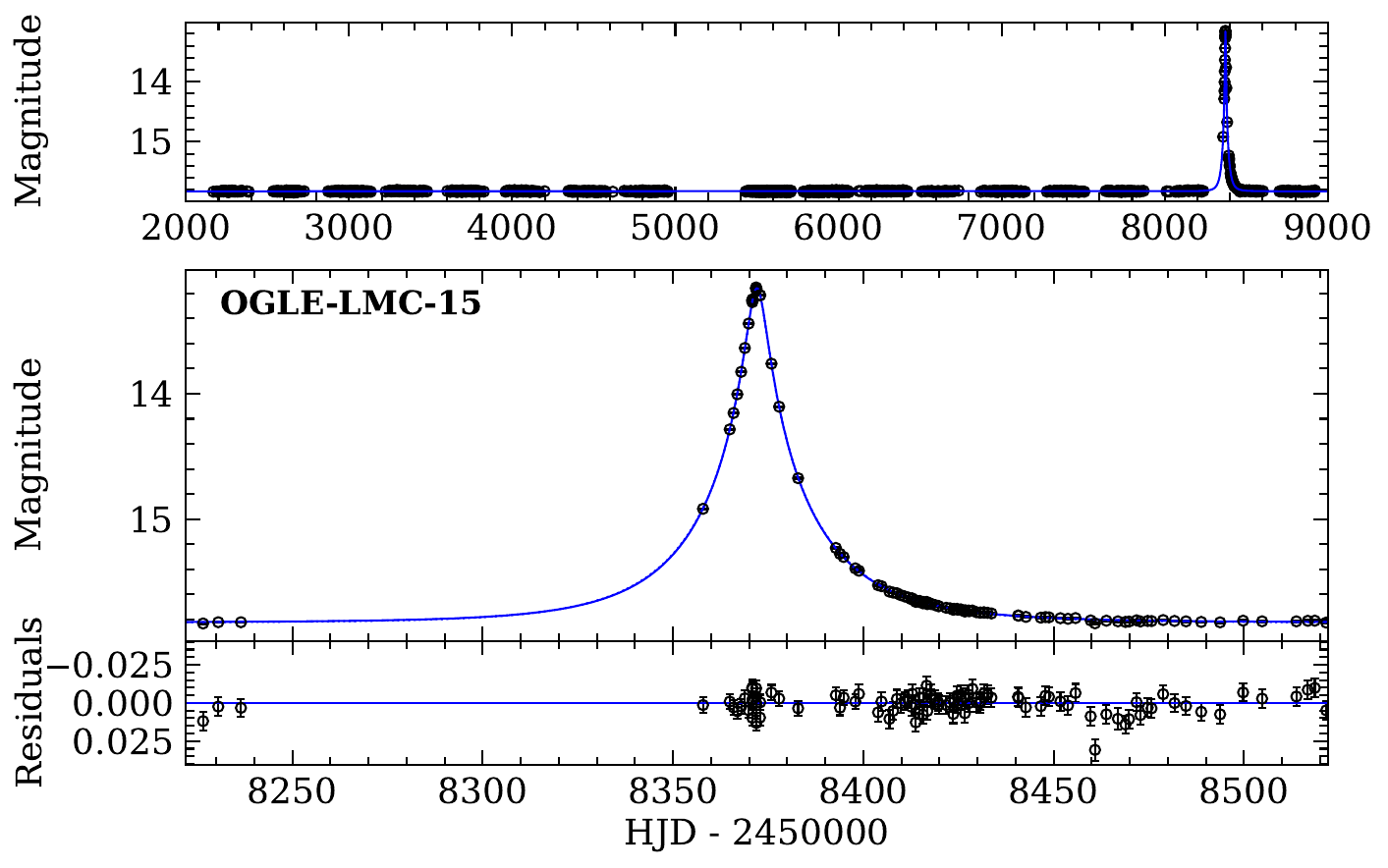}
\includegraphics[width=0.49\textwidth]{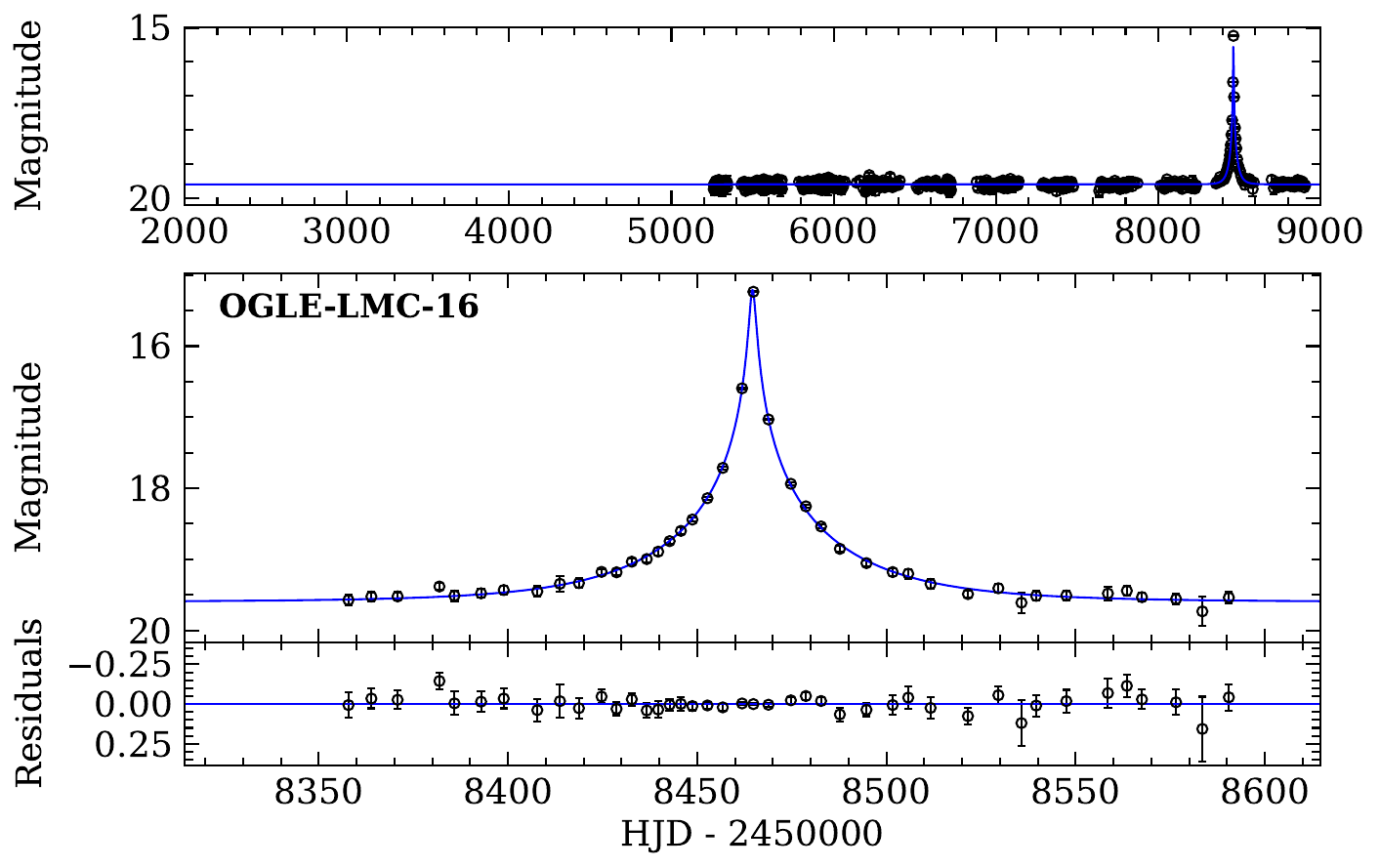}
\includegraphics[width=0.49\textwidth]{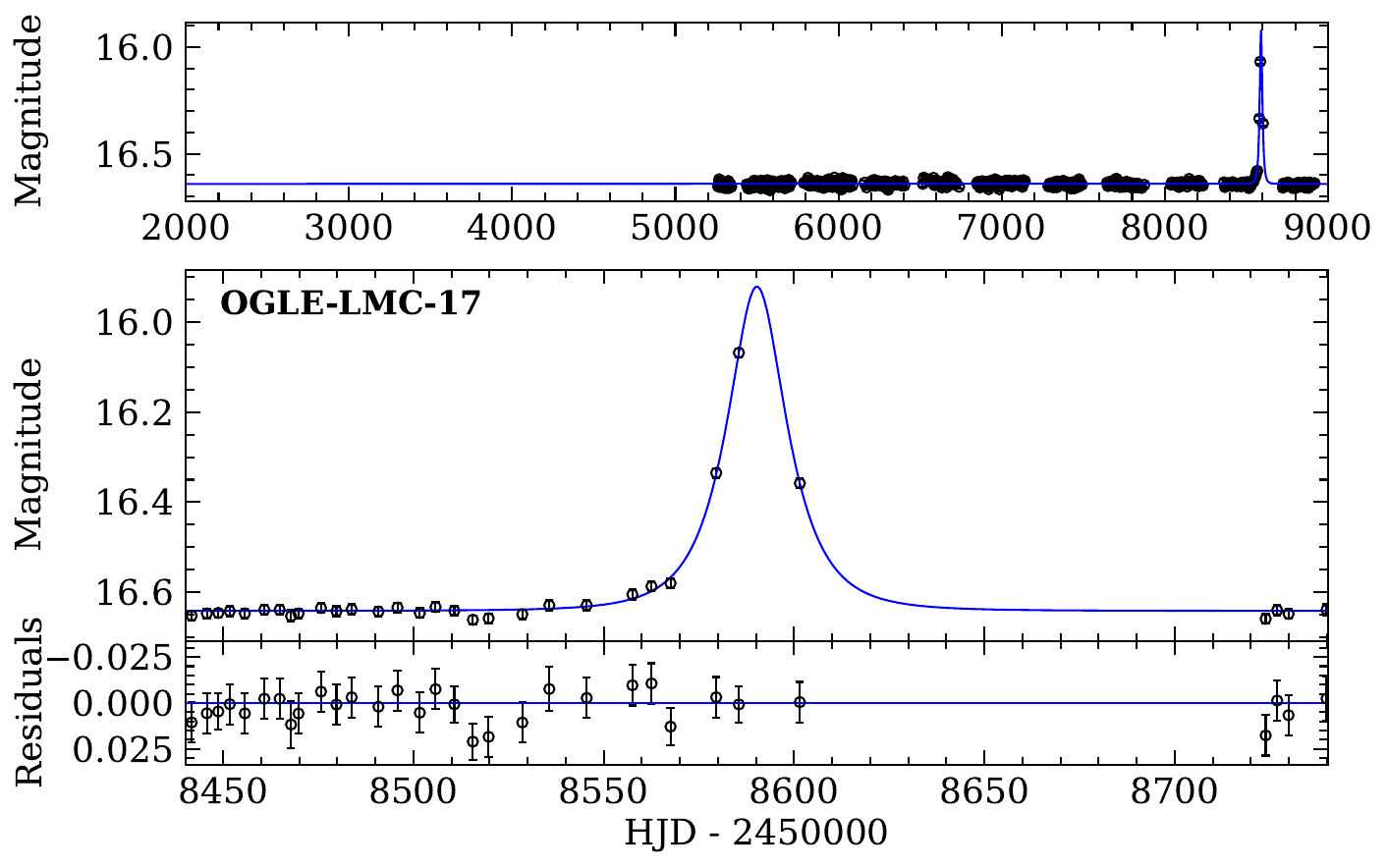}
\includegraphics[width=0.49\textwidth]{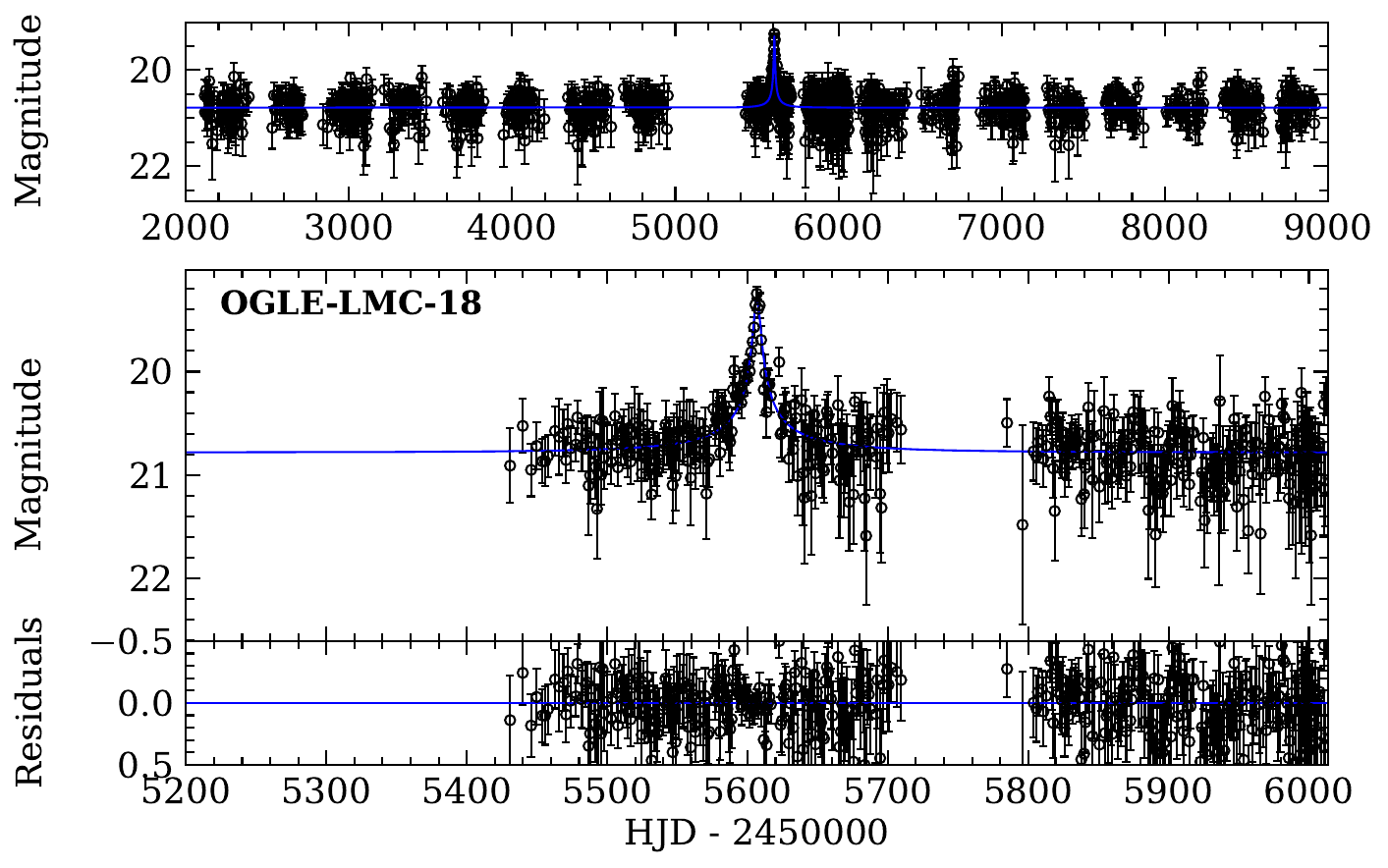}
\includegraphics[width=0.49\textwidth]{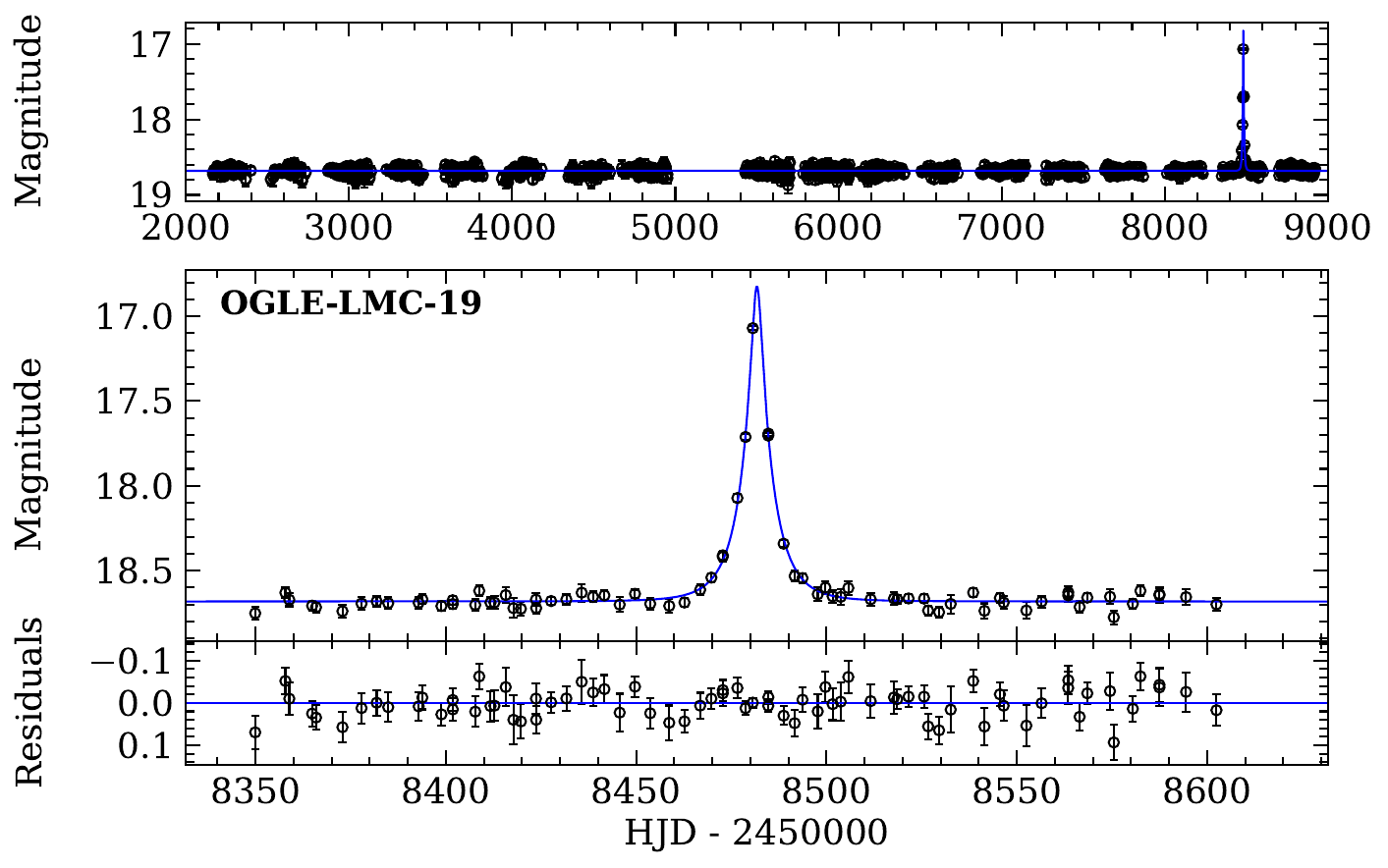}
\caption{(continuation)}
\end{figure*}

\begin{figure*}
\centering
\includegraphics[width=.5\textwidth]{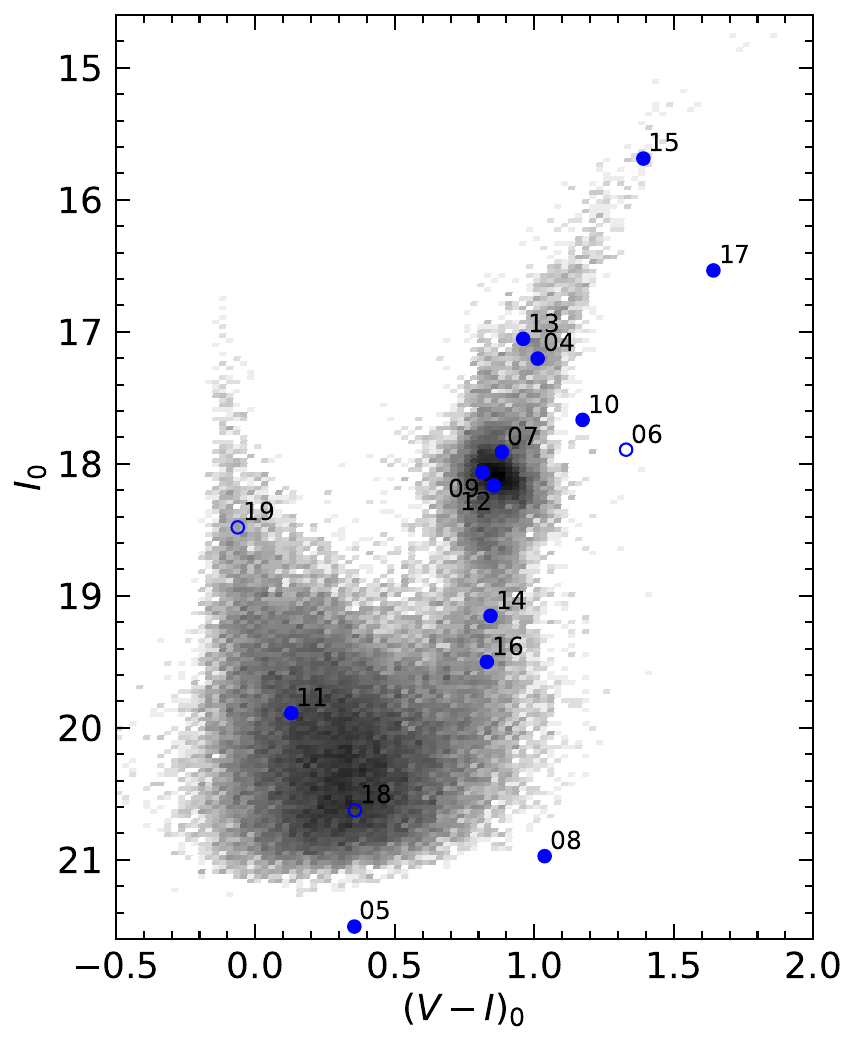}
\caption{Dereddened color--magnitude diagram of the field LMC516.14. Blue circles mark the positions of the detected microlensing events in the baseline. Empty circles mark three events that do not enter the final statistical sample.}
\label{fig:cmd}
\end{figure*}

\begin{figure*}
\centering
\includegraphics[width=.5\textwidth]{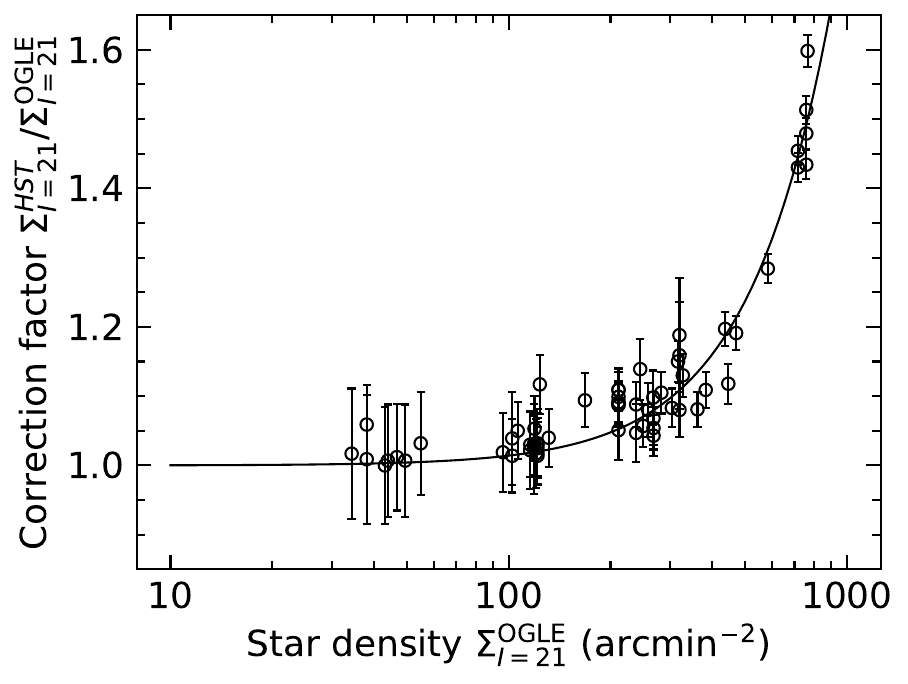}
\caption{Ratio between the mean surface density of stars brighter than $I=21$ detected in the HST and OGLE images (correction factor) as a function of density of stars observed by OGLE.}
\label{fig:corr}
\end{figure*}

\begin{figure*}
\centering
\includegraphics[width=.8\textwidth]{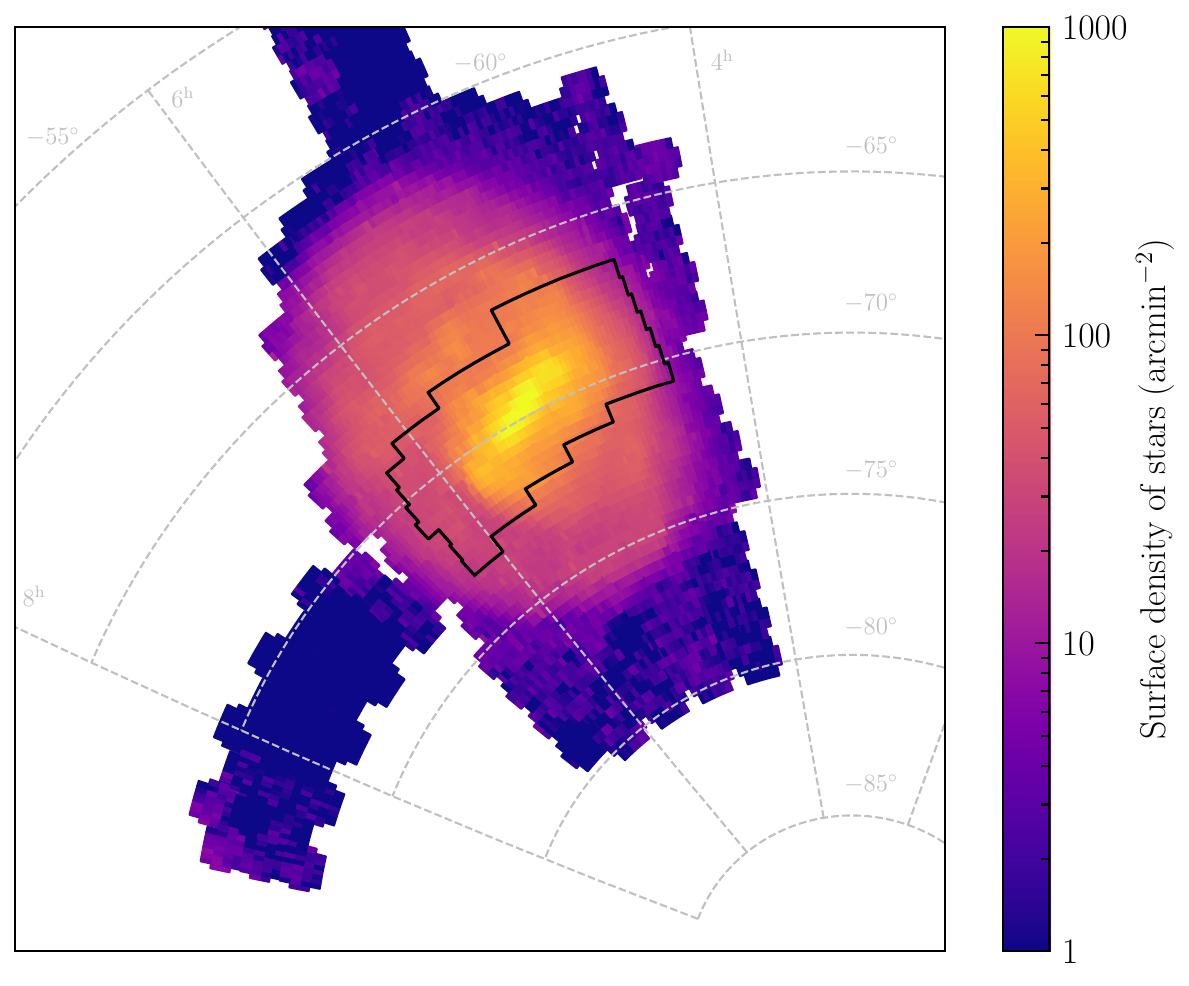}
\caption{Surface density of source stars brighter than $I=22$ in the analyzed OGLE-IV fields. The black polygon marks the region observed during the OGLE-III phase.}
\label{fig:stars}
\end{figure*}

\begin{figure}
\centering
\includegraphics[width=.5\textwidth]{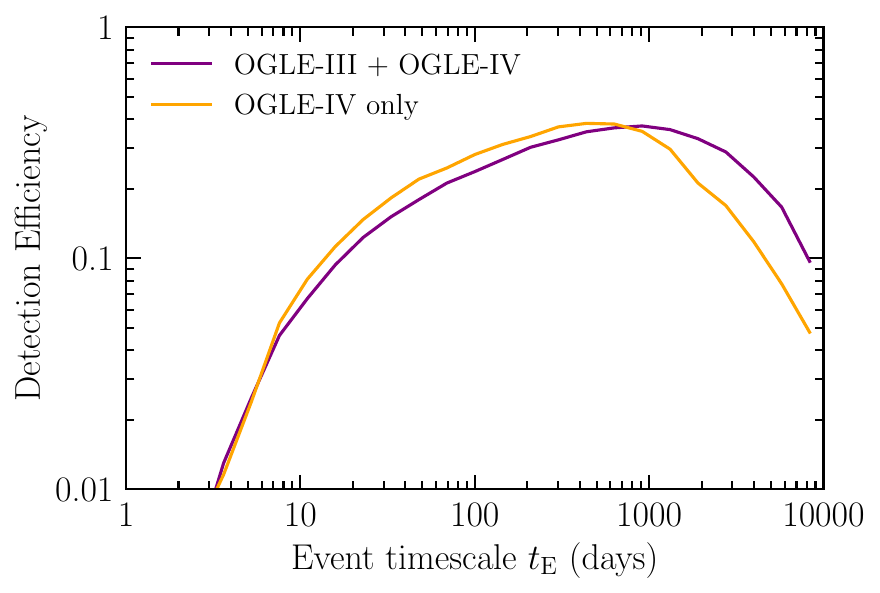}
\caption{Detection efficiency to microlensing events as a function of the Einstein timescale for field LMC501 (averaged for sources brighter than $I=22$ mag). The purple line marks the detection efficiency for stars observed during OGLE-III and OGLE-IV phases of the project, and the orange line -- for stars observed during OGLE-IV only.}
\label{fig:eff}
\end{figure}

\begin{figure*}
\centering
\includegraphics[width=.7\textwidth]{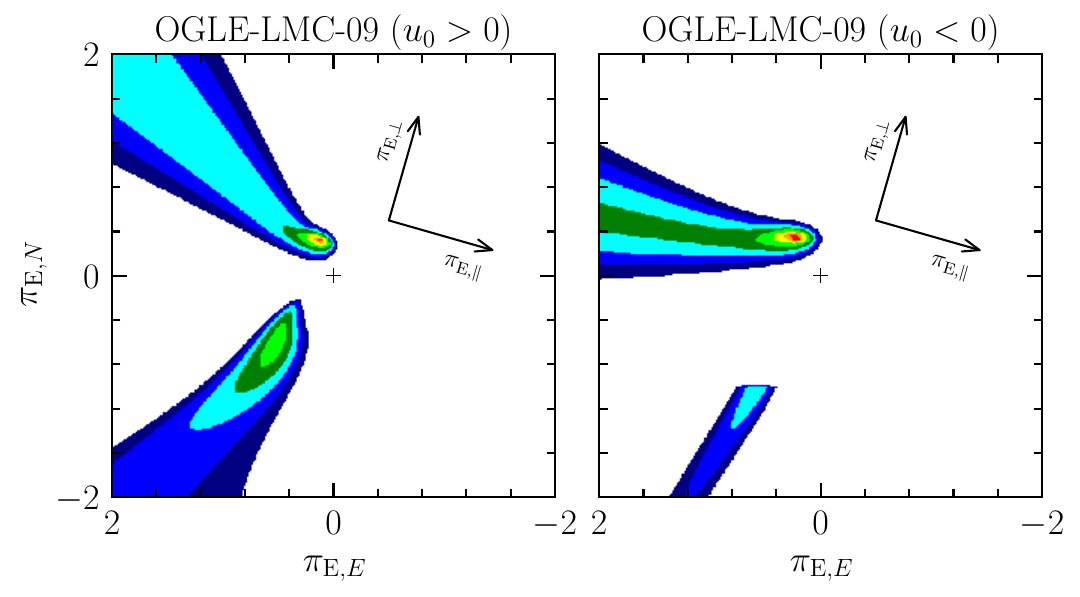}
\includegraphics[width=.7\textwidth]{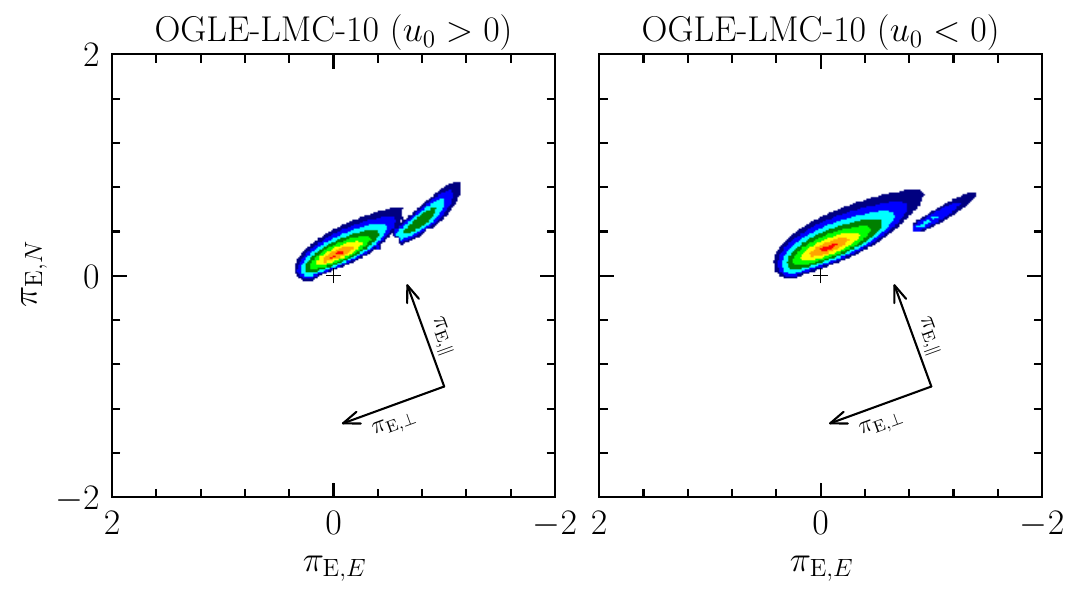}
\includegraphics[width=.7\textwidth]{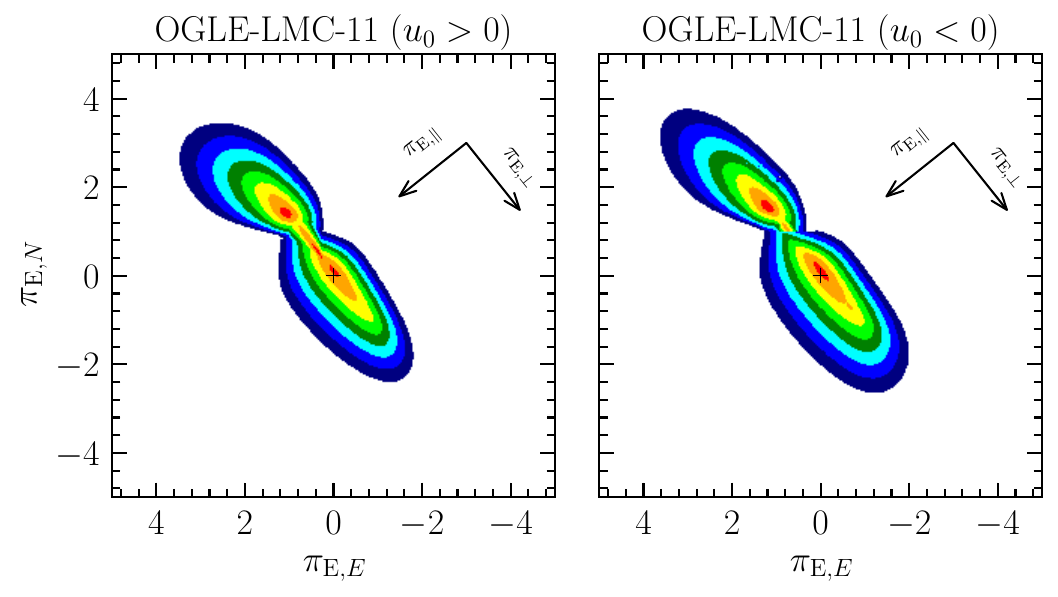}
\caption{Likelihood contours in the $(\piEN,\piEE)$ plane. The red, orange, yellow, lime, green, cyan, blue, and dark blue colors mark $\Delta\chi^2=1,4,9,16,25,36,49,64$, respectively. The arrows mark the directions of positive $\pi_{\rm E,\parallel}$ and $\pi_{\rm E,\perp}$, that is, directions parallel and perpendicular to the projected acceleration of the Sun in the adopted geocentric frames.}
\label{fig:parallaxes}
\end{figure*}

\addtocounter{figure}{-1}

\begin{figure*}
\centering
\includegraphics[width=.7\textwidth]{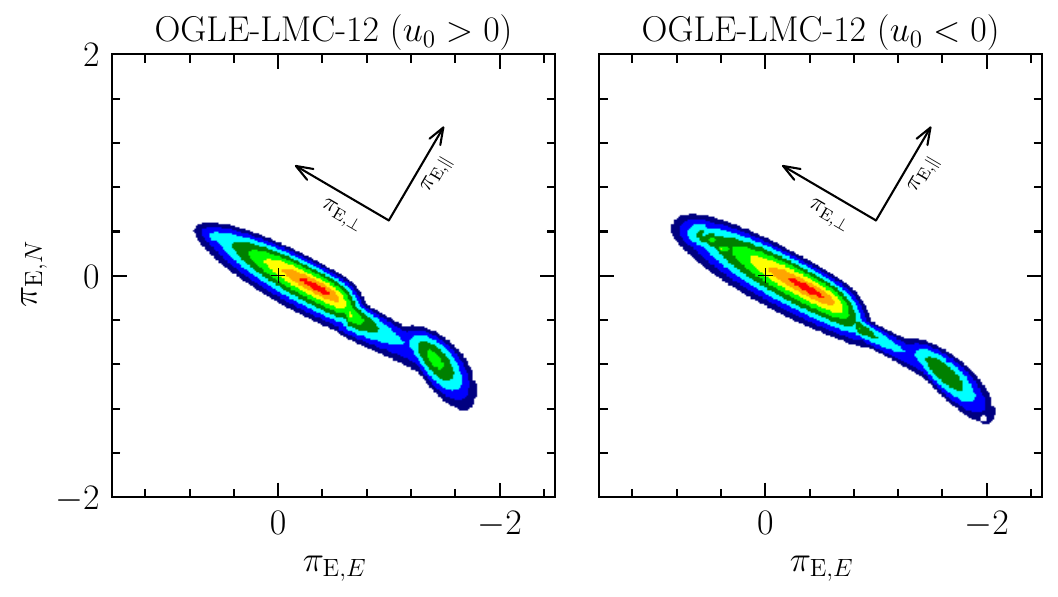}
\includegraphics[width=.7\textwidth]{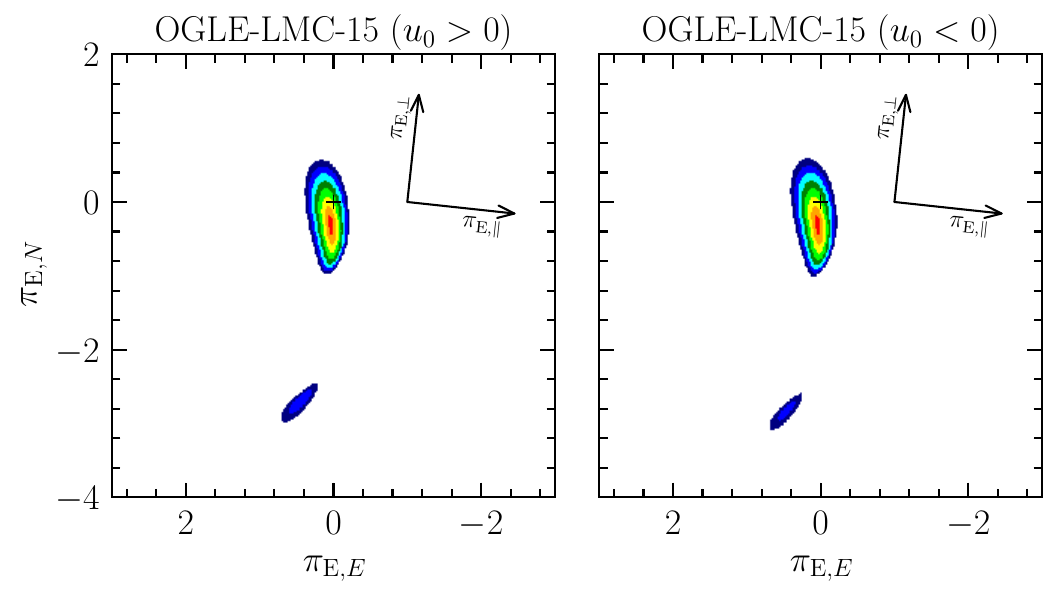}
\includegraphics[width=.7\textwidth]{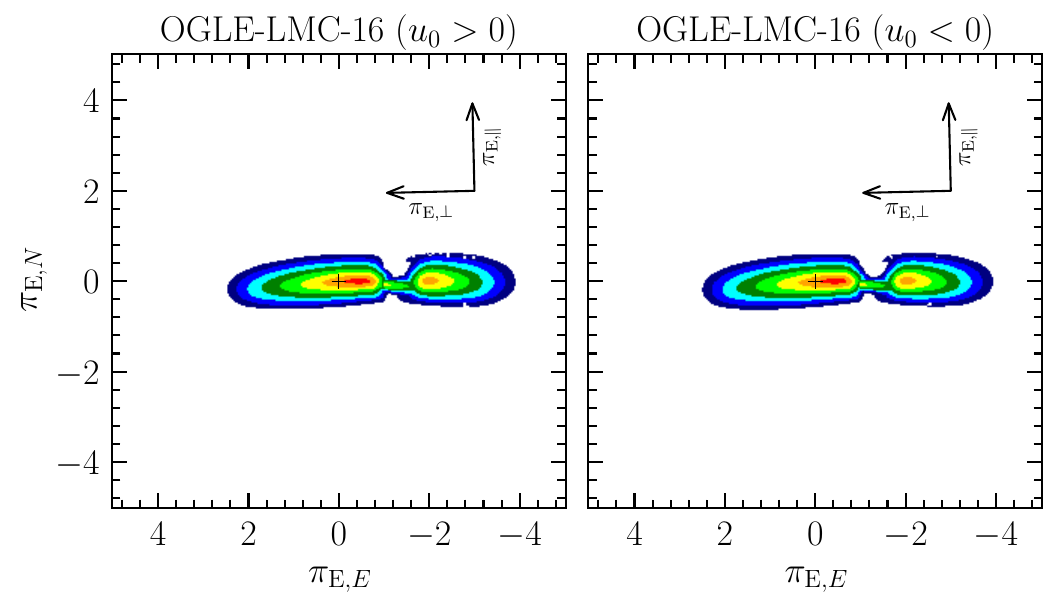}
\caption{(continuation)}
\end{figure*}

\clearpage

\begin{deluxetable*}{rrrrrrrr}
\tablecaption{OGLE-IV Fields toward the Large Magellanic Cloud\label{tab:fields}}
\tablehead{
\colhead{Field} & \colhead{R.A.} & \colhead{Decl.} & \colhead{$N_{\rm Epochs}$} & \colhead{$N_{\rm Epochs}$} & \colhead{$N_{\rm Stars}/1000$} & \colhead{$N_{\rm Stars}/1000$} & \colhead{$N_{\rm Stars}/1000$}\\
\colhead{} & \colhead{(J2000)} & \colhead{(J2000)} & \colhead{(2001--2009)} & \colhead{(2010--2020)} & \colhead{(2001--2020)} & \colhead{(2010--2020)} & \colhead{(Total)}
}
\startdata
LMC500 & $05^{\rm h}19^{\rm m}00^{\rm s}$ & $-73^{\circ}00{'}00{''}$ & 0 & 571 & 0.0 & 407.5 & 407.5 \\
LMC501 & $05^{\rm h}19^{\rm m}00^{\rm s}$ & $-71^{\circ}46{'}10{''}$ & 0--1779 & 559 & 582.5 & 205.3 & 787.8 \\
LMC502 & $05^{\rm h}19^{\rm m}00^{\rm s}$ & $-70^{\circ}32{'}20{''}$ & 445--2110 & 881 & 1833.1 & 0.0 & 1833.1 \\
LMC503 & $05^{\rm h}19^{\rm m}00^{\rm s}$ & $-69^{\circ}18{'}30{''}$ & 460--2191 & 884 & 3364.6 & 0.0 & 3364.6 \\
LMC504 & $05^{\rm h}19^{\rm m}00^{\rm s}$ & $-68^{\circ}04{'}40{''}$ & 0--1228 & 833 & 1150.8 & 149.0 & 1299.8 \\
LMC505 & $05^{\rm h}19^{\rm m}00^{\rm s}$ & $-66^{\circ}50{'}50{''}$ & 0--500 & 552 & 215.0 & 604.0 & 818.9 \\
LMC506 & $05^{\rm h}19^{\rm m}00^{\rm s}$ & $-65^{\circ}37{'}00{''}$ & 0 & 554 & 0.0 & 589.9 & 589.9 \\
LMC507 & $05^{\rm h}03^{\rm m}20^{\rm s}$ & $-72^{\circ}23{'}05{''}$ & 0 & 568 & 0.0 & 520.9 & 520.9 \\
LMC508 & $05^{\rm h}04^{\rm m}17^{\rm s}$ & $-71^{\circ}09{'}15{''}$ & 0--1769 & 806 & 796.3 & 158.3 & 954.6 \\
LMC509 & $05^{\rm h}05^{\rm m}07^{\rm s}$ & $-69^{\circ}55{'}25{''}$ & 427--1946 & 796 & 2041.3 & 0.0 & 2041.3 \\
\multicolumn{1}{c}{\dots} & \multicolumn{1}{c}{\dots} & \multicolumn{1}{c}{\dots} & \multicolumn{1}{c}{\dots} & \multicolumn{1}{c}{\dots} & \multicolumn{1}{c}{\dots} & \multicolumn{1}{c}{\dots} & \multicolumn{1}{c}{\dots}\\
Total  & \multicolumn{1}{c}{\dots} & \multicolumn{1}{c}{\dots} & \multicolumn{1}{c}{\dots} & \multicolumn{1}{c}{\dots} & 32969.9 & 29271.6 & 62241.5 \\
\enddata
\tablecomments{The coordinates are given for the epoch J2000. $N_{\rm Epochs}$ is the number of epochs, separately for the OGLE-III (2001--2009) and OGLE-IV (2010--2020) phases. {$N_{\rm Stars}$} is the number of stars (in thousands) observed in a given field; this is not equivalent to the number of microlensing source stars that enter the calculations (see the text). We discriminate between stars observed during both OGLE-III and OGLE-IV phases (with the available photometric data spanning 2001--2020) and those observed during OGLE-IV only (2010--2020). (This table is available in its entirety in machine-readable form.)}
\end{deluxetable*}

\clearpage

\begin{deluxetable*}{RRRRRR}
\tablecaption{Median Difference between the New OGLE-III and OGLE-IV Magnitudes\label{tab:color_shift}}
\tablehead{
\colhead{$V-I$} & \colhead{$\Delta I$} & \colhead{$V-I$} & \colhead{$\Delta I$} & \colhead{$V-I$} & \colhead{$\Delta I$}}
\startdata
-0.171 & -0.068 & 0.876 & -0.020 & 1.624 & 0.020 \\
-0.123 & -0.064 & 0.927 & -0.016 & 1.674 & 0.022 \\
-0.075 & -0.061 & 0.976 & -0.014 & 1.722 & 0.024 \\
-0.027 & -0.057 & 1.025 & -0.011 & 1.772 & 0.027 \\
0.023 & -0.053 & 1.076 & -0.008 & 1.823 & 0.029 \\
0.095 & -0.044 & 1.127 & -0.005 & 1.873 & 0.031 \\
0.196 & -0.034 & 1.177 & -0.002 & 1.939 & 0.034 \\
0.298 & -0.030 & 1.228 & 0.001 & 2.044 & 0.037 \\
0.402 & -0.028 & 1.277 & 0.004 & 2.145 & 0.040 \\
0.500 & -0.028 & 1.326 & 0.006 & 2.242 & 0.047 \\
0.608 & -0.029 & 1.375 & 0.008 & 2.350 & 0.053 \\
0.680 & -0.030 & 1.425 & 0.010 & 2.440 & 0.059 \\
0.730 & -0.028 & 1.474 & 0.013 & 2.602 & 0.072 \\
0.776 & -0.026 & 1.524 & 0.015 &  &  \\
0.825 & -0.023 & 1.574 & 0.017 &  &  \\
\enddata
\end{deluxetable*}

\begin{deluxetable*}{rcccc}
\tablecaption{Error Bar Correction Coefficients for the $I$-band Data\label{tab:errors}}
\tablehead{
\colhead{Field} & \colhead{$\gamma_{\operatorname{OGLE-III}}$} & \colhead{$\varepsilon_{\operatorname{OGLE-III}}$} & \colhead{$\gamma_{\operatorname{OGLE-IV}}$} & \colhead{$\varepsilon_{\operatorname{OGLE-IV}}$}}
\startdata
LMC501.01 & \dots & \dots  & 1.115 & 0.0062 \\
LMC501.02 & \dots & \dots  & 1.069 & 0.0057 \\
LMC501.03 & \dots & \dots  & 1.155 & 0.0063 \\
LMC501.04 & \dots & \dots  & 1.178 & 0.0059 \\
LMC501.05 & \dots & \dots  & 1.182 & 0.0057 \\
LMC501.06 & \dots & \dots  & 1.157 & 0.0061 \\
LMC501.07 & \dots & \dots  & 1.156 & 0.0063 \\
LMC501.08 & 1.189 & 0.0063 & 1.127 & 0.0043 \\
LMC501.09 & 1.206 & 0.0062 & 1.087 & 0.0042 \\
LMC501.10 & 1.191 & 0.0081 & 1.082 & 0.0053 \\
\multicolumn{1}{c}{\dots} & \multicolumn{1}{c}{\dots} & \multicolumn{1}{c}{\dots} & \multicolumn{1}{c}{\dots} & \multicolumn{1}{c}{\dots} \\
\enddata
\tablecomments{(This table is available in its entirety in machine-readable form.)}
\end{deluxetable*}

\begin{table*}
\centering
\footnotesize
\caption{Selection Cuts}
\label{tab:cuts}
\begin{tabular}
    {p{0.1\textwidth}%
    p{0.6\textwidth}%
    >{\raggedleft\arraybackslash}p{0.1\textwidth}}
\hline
\textbf{Cut 0.} & \textbf{All Stars in the Databases} & 62,591,045 \\
\hline
\textbf{Cut 1.} & \textbf{Stars with at least one significant brightening in the light curve} \\
& At least five consecutive data points $3\sigma$ above the baseline flux ($n_{\rm bump} \geq 5$) & \\
& Object detected on at least three subtracted images ($n_{\rm DIA} \geq 3$) &\\
& The total ``significance'' of the bump ($\chi_{3+}=\sum_i(F_i-F_{\rm base})/\sigma_i \geq 32$) &\\
& Amplitude of the bump at least 0.1 mag ($\Delta m \geq 0.1$ mag) & \\
& Amplitude of the bump at least 0.4 mag if the bump is longer than 100 days & \\
& If a bump is shorter than 1000 days, no significant variability outside the window ($\chi^2_{\rm out}/\mathrm{d.o.f.} \leq 2$) & 56,358 \\
\hline
\textbf{Cut 2.} & \textbf{Removing false positives} & \\
& Stars with multiple bumps in the data & \\
& ``Blue bumpers'' ($(V-I)_0 \leq 0.5$, $I_0 \leq 19.5$) & \\
& Stars in the vicinity of SN 1987A & 49,325 \\
\hline
\textbf{Cut 3.} & \textbf{Microlensing model describes the light curve well} & \\
& Fit converged & \\
& $\chi^2/\mathrm{d.o.f.} \leq 2$ (all data points) & \\
& $\chi^2_{\tE}/\mathrm{d.o.f.} \leq 2$ (for $|t - t_0| < \tE$) & \\
& $\chi^2_{\rm bump}/\mathrm{d.o.f.} \leq 2$ (for data points within the bump) & \\
& $t_0$ within the time range covered by the data & \\
& Impact parameter is smaller than 1 ($u_0 \leq 1$) & \\ 
& Uncertainty on the Einstein timescale ($\sigma(\tE) / \tE \leq 1$) & \\
& Source star brighter than 22 mag ($I_{\rm s} \leq 22$) & \\
& Microlensing model is significantly better than a straight line (\mbox{$\chi^2_{\rm line} - \chi^2 \geq 250 \chi^2/\mathrm{d.o.f.}$}) & 13 \\
\hline
\end{tabular}
\end{table*}

\begin{deluxetable*}{ccclcc}
\centering
\tablecaption{Detected Microlensing Events\label{tab:events}}
\tablehead{
\colhead{Event} & \colhead{R.A. (J2000)} & \colhead{Decl. (J2000)} & \colhead{OGLE-IV ID}& \colhead{$I$ (mag)} & \colhead{$V-I$ (mag)}}
\startdata
OGLE-LMC-04 & \ra{05}{25}{39}{57} & \dec{-70}{19}{49}{6} & LMC502.17.57865 & 17.31 & 1.10 \\
OGLE-LMC-05 & \ra{05}{24}{49}{10} & \dec{-67}{50}{04}{9} & LMC504.17.27728 & 21.67 & 0.49 \\
OGLE-LMC-06 & \ra{05}{19}{47}{80} & \dec{-70}{46}{26}{6} & LMC502.12.400   & 18.04 & 1.45 \\
OGLE-LMC-07 & \ra{05}{21}{07}{64} & \dec{-69}{34}{45}{1} & LMC503.11.3147  & 17.99 & 0.95 \\
OGLE-LMC-08 & \ra{05}{12}{29}{58} & \dec{-68}{05}{50}{6} & LMC504.16.27046 & 21.11 & 1.15 \\
OGLE-LMC-09 & \ra{05}{00}{23}{51} & \dec{-69}{30}{49}{9} & LMC509.32.3044  & 18.23 & 0.95 \\
OGLE-LMC-10 & \ra{05}{41}{47}{97} & \dec{-70}{40}{52}{1} & LMC552.15.15750 & 17.98 & 1.43 \\
OGLE-LMC-11 & \ra{05}{14}{10}{19} & \dec{-67}{49}{14}{0} & LMC504.24.15713 & 20.00 & 0.22 \\
OGLE-LMC-12 & \ra{05}{16}{31}{35} & \dec{-71}{05}{11}{8} & LMC502.05.27026 & 18.28 & 0.95 \\
OGLE-LMC-13 & \ra{05}{22}{11}{57} & \dec{-69}{33}{29}{5} & LMC503.10.65388 & 17.09 & 0.99 \\
OGLE-LMC-14 & \ra{05}{14}{33}{55} & \dec{-68}{50}{58}{8} & LMC503.32.13953 & 19.27 & 0.94 \\
OGLE-LMC-15 & \ra{05}{33}{05}{32} & \dec{-69}{30}{31}{2} & LMC516.29.8037  & 15.82 & 1.50 \\
OGLE-LMC-16 & \ra{05}{07}{45}{58} & \dec{-65}{57}{49}{9} & LMC512.21.20190 & 19.61 & 0.92 \\
OGLE-LMC-17 & \ra{06}{07}{25}{55} & \dec{-67}{19}{41}{1} & LMC570.03.200   & 16.63 & 1.72 \\
OGLE-LMC-18 & \ra{05}{21}{25}{25} & \dec{-69}{12}{54}{7} & LMC503.20.21374 & 20.75 & 0.46 \\
OGLE-LMC-19 & \ra{04}{46}{14}{25} & \dec{-69}{03}{20}{2} & LMC531.25.9430  & 18.68 & 0.10 \\
\enddata
\tablecomments{The table provides the mean $I$-band brightness and $V-I$ color in the baseline.}
\end{deluxetable*}


\movetabledown=2in
\begin{rotatetable}
\begin{deluxetable*}{lrrrrrrrrr}
\tabletypesize{\scriptsize}
\tablecaption{Best-fit Parameters of Detected Events\label{tab:params}}
\tablehead{
\colhead{Event} & \colhead{$t_0$} & \colhead{$t_{\rm E}$ (d)} & \colhead{$u_0$} & \colhead{$\piEN$} & \colhead{$\piEE$} & \colhead{$I_{\rm s}$ (mag)} & \colhead{$f_{\rm s}$} & \colhead{$\chi^2$} & \colhead{$N$}}
\startdata
OGLE-LMC-04 & $2452229.045 \pm 0.719$ & $35.165^{+8.302}_{-4.622}$ & $0.825^{+0.199}_{-0.231}$ & \dots & \dots & $17.929^{+0.626}_{-0.467}$ & $0.575^{+0.309}_{-0.252}$ & 638.1 & 581\\
OGLE-LMC-05 & $2453104.123^{+4.389}_{-3.857}$ & $165.941^{+87.842}_{-49.125}$ & $0.219^{+0.162}_{-0.099}$ & \dots & \dots & $21.428 \pm 0.765$ & $1.196^{+1.202}_{-0.583}$ & 308.0 & 436\\
OGLE-LMC-07 & $2452269.233 \pm 1.444$ & $52.127^{+72.009}_{-25.366}$ & $0.362^{+0.622}_{-0.246}$ & \dots & \dots & $20.646^{+1.480}_{-1.798}$ & $0.086^{+0.366}_{-0.064}$ & 428.4 & 500\\
OGLE-LMC-08 & $2454856.318^{+0.073}_{-0.081}$ & $13.497^{+6.005}_{-3.924}$ & $0.053^{+0.035}_{-0.024}$ & \dots & \dots & $21.093 \pm 0.509$ & $0.990^{+0.579}_{-0.369}$ & 432.4 & 467\\
OGLE-LMC-09 & $2455432.621 \pm 0.469$ & $70.383 \pm 1.736$ & $0.387 \pm 0.018$ & \dots & \dots & $17.772 \pm 0.068$ & $1.623 \pm 0.104$ & 997.6 & 818\\
OGLE-LMC-09 & $2455432.025 \pm 0.811$ & $102.756^{+23.386}_{-17.555}$ & $0.181^{+0.057}_{-0.046}$ & $0.315 \pm 0.018$ & $0.132^{+0.034}_{-0.028}$ & $18.785 \pm 0.368$ & $0.636^{+0.252}_{-0.183}$ & 745.3 & 818\\
OGLE-LMC-09 & $2455431.551 \pm 1.377$ & $120.063^{+36.393}_{-27.518}$ & $-0.184^{+0.052}_{-0.075}$ & $0.343 \pm 0.018$ & $0.242^{+0.053}_{-0.042}$ & $18.757 \pm 0.455$ & $0.652^{+0.340}_{-0.209}$ & 745.3 & 818\\
OGLE-LMC-10 & $2455568.724 \pm 0.160$ & $80.791 \pm 4.204$ & $0.247 \pm 0.019$ & \dots & \dots & $18.971 \pm 0.096$ & $0.453^{+0.042}_{-0.038}$ & 620.6 & 587\\
OGLE-LMC-10 & $2455570.014 \pm 0.201$ & $64.595^{+6.134}_{-5.008}$ & $0.340 \pm 0.048$ & $0.197 \pm 0.039$ & $-0.040 \pm 0.075$ & $18.520 \pm 0.200$ & $0.686^{+0.134}_{-0.116}$ & 491.6 & 587\\
OGLE-LMC-10 & $2455569.832 \pm 0.225$ & $61.767^{+8.168}_{-6.581}$ & $-0.337 \pm 0.045$ & $0.248 \pm 0.045$ & $-0.057 \pm 0.096$ & $18.533 \pm 0.205$ & $0.678^{+0.137}_{-0.116}$ & 491.9 & 587\\
OGLE-LMC-11 & $2455670.398 \pm 0.089$ & $41.474^{+3.122}_{-2.771}$ & $0.148 \pm 0.015$ & \dots & \dots & $19.932 \pm 0.118$ & $1.178 \pm 0.130$ & 828.0 & 824\\
OGLE-LMC-12 & $2456974.451 \pm 0.099$ & $45.923 \pm 1.517$ & $0.216 \pm 0.011$ & \dots & \dots & $18.329 \pm 0.063$ & $0.962 \pm 0.057$ & 874.1 & 877\\
OGLE-LMC-12 & $2456974.623 \pm 0.124$ & $45.051^{+2.033}_{-1.716}$ & $0.227 \pm 0.016$ & $-0.089^{+0.077}_{-0.069}$ & $-0.306^{+0.144}_{-0.123}$ & $18.264^{+0.093}_{-0.083}$ & $1.022 \pm 0.084$ & 865.9 & 877\\
OGLE-LMC-12 & $2456974.570 \pm 0.115$ & $42.890^{+3.076}_{-2.400}$ & $-0.229 \pm 0.019$ & $-0.100^{+0.097}_{-0.085}$ & $-0.336^{+0.174}_{-0.149}$ & $18.254^{+0.107}_{-0.096}$ & $1.031 \pm 0.097$ & 865.9 & 877\\
OGLE-LMC-13 & $2457138.949 \pm 0.192$ & $7.007^{+1.936}_{-1.059}$ & $0.638 \pm 0.203$ & \dots & \dots & $17.730^{+0.636}_{-0.490}$ & $0.553^{+0.315}_{-0.245}$ & 784.4 & 861\\
OGLE-LMC-14 & $2457759.472 \pm 3.487$ & $88.188^{+59.008}_{-25.697}$ & $0.581^{+0.399}_{-0.310}$ & \dots & \dots & $20.285^{+1.156}_{-1.023}$ & $0.392^{+0.614}_{-0.257}$ & 931.4 & 868\\
OGLE-LMC-15 & $2458372.030 \pm 0.009$ & $32.204 \pm 0.185$ & $0.081 \pm 0.001$ & \dots & \dots & $15.892 \pm 0.010$ & $0.941 \pm 0.008$ & 1047.7 & 899\\
OGLE-LMC-15 & $2458372.034 \pm 0.010$ & $31.278 \pm 0.459$ & $0.085 \pm 0.002$ & $-0.317 \pm 0.135$ & $0.047 \pm 0.036$ & $15.845 \pm 0.021$ & $0.983 \pm 0.019$ & 1036.0 & 899\\
OGLE-LMC-15 & $2458372.033 \pm 0.010$ & $31.000 \pm 0.463$ & $-0.085 \pm 0.002$ & $-0.320 \pm 0.136$ & $0.051 \pm 0.040$ & $15.843 \pm 0.021$ & $0.984 \pm 0.019$ & 1036.0 & 899\\
OGLE-LMC-16 & $2458464.623 \pm 0.024$ & $41.800 \pm 1.599$ & $0.019 \pm 0.001$ & \dots & \dots & $19.497 \pm 0.049$ & $1.100 \pm 0.050$ & 497.5 & 591\\
OGLE-LMC-17 & $2458590.262 \pm 0.225$ & $13.790^{+1.836}_{-1.046}$ & $0.491^{+0.075}_{-0.106}$ & \dots & \dots & $16.898^{+0.336}_{-0.219}$ & $0.790^{+0.177}_{-0.210}$ & 521.2 & 657\\
OGLE-LMC-18 & $2455606.738 \pm 0.240$ & $101.230^{+89.561}_{-38.668}$ & $0.026^{+0.019}_{-0.013}$ & \dots & \dots & $23.572^{+0.752}_{-0.611}$ & $0.076^{+0.057}_{-0.038}$ & 903.8 & 868\\
OGLE-LMC-19 & $2458481.729 \pm 0.038$ & $8.200^{+0.690}_{-0.599}$ & $0.168 \pm 0.026$ & \dots & \dots & $18.799 \pm 0.146$ & $0.898 \pm 0.125$ & 1261.8 & 1227\\
\enddata
\end{deluxetable*}
\end{rotatetable}

%
\begin{deluxetable*}{rc}
\tablecaption{Color of the Source Star from the Model-independent Regression\label{tab:colors}}
\tablehead{
\colhead{Event} & \colhead{$(V-I)_{\rm s}$}}
\startdata
OGLE-LMC-05 & $0.51 \pm 0.22$\tablenotemark{a}\\
OGLE-LMC-09 & $1.011 \pm 0.020$ \\
OGLE-LMC-10 & $1.285 \pm 0.047$ \\
OGLE-LMC-11 & $0.221 \pm 0.020$ \\
OGLE-LMC-12 & $0.906 \pm 0.036$ \\
OGLE-LMC-13 & $0.906 \pm 0.048$ \\
OGLE-LMC-15 & $1.476 \pm 0.004$ \\
\enddata
\tablenotetext{a}{\citet{wyrzykowski2011}}
\end{deluxetable*}

\begin{deluxetable*}{rrrrrrrrr}
\tablecaption{Number of Microlensing Source Stars Observed in OGLE-IV Fields\label{tab:stars}}
\tablehead{
\colhead{Field} & \colhead{R.A. (deg)} & \colhead{Decl. (deg)} & \colhead{$N_{\rm total}^{\rm 21\,mag}$} & \colhead{$N_{\rm corr}^{\rm 21\,mag}$} & \colhead{$N_{\rm total}^{\rm 21.5\,mag}$} & \colhead{$N_{\rm corr}^{\rm 21.5\,mag}$} & \colhead{$N_{\rm total}^{\rm 22\,mag}$} & \colhead{$N_{\rm corr}^{\rm 22\,mag}$}}
\startdata
LMC500.01 & 81.3326 & --73.4710 &  8119 &  7106 & 11055 &  9977 & 12185 & 11055 \\
LMC500.02 & 80.8051 & --73.4710 &  8193 &  7187 & 10559 &  9489 & 11558 & 10437 \\
LMC500.03 & 80.2775 & --73.4710 &  7780 &  6780 &  9527 &  8463 & 10346 &  9232 \\
LMC500.04 & 79.7500 & --73.4710 &  7412 &  6420 &  9859 &  8803 & 10752 &  9645 \\
LMC500.05 & 79.2225 & --73.4710 &  7565 &  6579 & 10020 &  8970 & 10933 &  9834 \\
LMC500.06 & 78.6949 & --73.4710 &  7098 &  6118 &  9760 &  8718 & 10656 &  9564 \\
LMC500.07 & 78.1674 & --73.4710 &  7846 &  6873 & 10624 &  9588 & 11662 & 10576 \\
LMC500.08 & 81.8602 & --73.1496 &  9457 &  8437 & 13042 & 11958 & 14538 & 13401 \\
LMC500.09 & 81.3326 & --73.1496 &  9892 &  8880 & 12295 & 11218 & 13586 & 12457 \\
LMC500.10 & 80.8051 & --73.1496 & 10252 &  9246 & 13021 & 11952 & 14471 & 13350 \\
\multicolumn{1}{c}{\dots} & \multicolumn{1}{c}{\dots} & \multicolumn{1}{c}{\dots} & \multicolumn{1}{c}{\dots} & \multicolumn{1}{c}{\dots} & \multicolumn{1}{c}{\dots} & \multicolumn{1}{c}{\dots} & \multicolumn{1}{c}{\dots} & \multicolumn{1}{c}{\dots}\\
Total (million) & \multicolumn{1}{c}{\dots} & \multicolumn{1}{c}{\dots} & 47.126 & 39.763 & 67.992 & 60.175 & 86.861 & 78.682\\
\enddata
\tablecomments{The coordinates are given for the epoch J2000. $N_{\rm total}$ is the number of possible microlensing source stars (after correcting the raw star counts for blending), and $N_{\rm corr}$ is the number of possible microlensing source stars after removing foreground Milky Way sources. (This table is available in its entirety in machine-readable form.)}
\end{deluxetable*}

\begin{deluxetable*}{lcccccc}
\tablecaption{Microlensing Optical Depth and Event Rate toward the LMC\label{tab:tau}}
\tablehead{
\colhead{Event} & \colhead{$\langle\varepsilon(\tE)\rangle$} & \colhead{$\langle1/\varepsilon(\tE)\rangle$} & \colhead{$\langle\tE/\varepsilon(\tE)\rangle$} & \colhead{$\Delta T$} & \colhead{$\tau$} & \colhead{$\Gamma$}\\
\colhead{} & \colhead{} & \colhead{} & \colhead{(days)} & \colhead{(days)} & \colhead{($10^{-7}$)} & \colhead{$(10^{-7}\,\mathrm{yr}^{-1})$}}
\startdata
OGLE-LMC-04 & 0.1671 &  6.018 &  219.32 & 7000 & 0.0063 & 0.0399 \\
OGLE-LMC-05 & 0.2692 &  3.770 &  688.01 & 7000 & 0.0196 & 0.0250 \\
OGLE-LMC-07 & 0.1749 &  6.219 &  444.43 & 7000 & 0.0127 & 0.0412 \\
OGLE-LMC-08 & 0.0826 & 13.338 &  175.12 & 7000 & 0.0050 & 0.0885 \\
OGLE-LMC-09 & 0.2467 &  4.076 &  425.42 & 7000 & 0.0121 & 0.0270 \\
OGLE-LMC-10 & 0.1911 &  5.240 &  340.83 & 7000 & 0.0097 & 0.0347 \\
OGLE-LMC-11 & 0.1652 &  6.086 &  252.41 & 7000 & 0.0072 & 0.0404 \\
OGLE-LMC-12 & 0.1872 &  5.343 &  241.59 & 7000 & 0.0069 & 0.0354 \\
OGLE-LMC-13 & 0.0419 & 25.364 &  179.82 & 7000 & 0.0051 & 0.1682 \\
OGLE-LMC-14 & 0.2090 &  4.886 &  483.57 & 7000 & 0.0138 & 0.0324 \\
OGLE-LMC-15 & 0.1334 &  7.494 &  234.51 & 7000 & 0.0067 & 0.0497 \\
OGLE-LMC-16 & 0.1960 &  5.114 &  213.74 & 4000 & 0.0107 & 0.0593 \\
OGLE-LMC-17 & 0.1261 &  8.139 &  113.76 & 4000 & 0.0057 & 0.0945 \\
\hline
\textbf{Total:} & \dots & \dots & \dots & \dots & $\mathbf{0.121 \pm 0.037}$ & $\mathbf{0.74 \pm 0.25}$\\
\enddata
\end{deluxetable*}

\begin{deluxetable*}{lccccc}
\tablecaption{Jerk-parallax Degeneracy in the Analyzed Events ($u_0 > 0$ Solutions)\label{tab:paral}}
\tablehead{
\colhead{Event} & \colhead{$(\piEN,\piEE)$} & \colhead{$(\piEN',\piEE')$} & \colhead{$(
\pi_{\rm E,\parallel}, \pi_{\rm E,\perp})$} & \colhead{$(\pi_{\rm E,\parallel}', \pi_{\rm E,\perp}')$} & \colhead{$(\pi_{j,\parallel}, \pi_{j,\perp})$}
}
\startdata
OGLE-LMC-09 & $(0.32,0.12)$  & $(-0.60,0.52)$ & $(-0.20,0.27)$ & $(-0.33,-0.72)$ & $(0.01,0.72)$ \\
OGLE-LMC-10 & $(0.20,-0.04)$ & $(0.48,-0.76)$ & $(0.17,-0.11)$ & $(0.19,-0.88)$  & $(-0.03,1.26)$ \\
OGLE-LMC-11 & $(0.08,0.00)$  & $(1.40,1.08)$  & $(-0.05,-0.06)$ & $(-0.03,-1.77)$ & $(-0.03,1.70)$ \\
OGLE-LMC-12 & $(-0.12,-0.36)$ & $(-0.76,-1.48)$ & $(0.08,-0.37)$ & $(0.09,-1.66)$ & $(0.13,1.69)$ \\
OGLE-LMC-15 & $(-0.32,0.04)$ & $(-2.68,0.52)$ & $(-0.01,-0.32)$ & $(-0.23,-2.72)$ & $(0.02,2.68)$ \\
OGLE-LMC-16 & $(0.00,-0.44)$ & $(0.00,-1.96)$ & $(-0.01,-0.44)$ & $(-0.05,-1.96)$ & $(-0.04,1.90)$
\enddata
\end{deluxetable*}

\end{document}